\documentclass[a4wide,useAMS,usenatbib]{mn2e}

\usepackage{graphicx}
\usepackage{natbib}
\usepackage{longtable}

\usepackage{txfonts}
\title[Herschel Virgo Cluster]{The Herschel Virgo Cluster Survey VIII: The Bright Galaxy Sample\thanks{{\it Herschel} is an ESA space observatory with science instruments provided by European-led Principal Investigator consortia and with important participation from NASA.}}
\author[Davies et al.]
{J. I. Davies$^{1}$,
S. Bianchi$^{2}$,
L. Cortese$^{5}$,
R. Auld$^{1}$,
M. Baes$^{3}$, 
G. J. Bendo$^{4}$,
A. Boselli$^{7}$,  \newauthor
L. Ciesla$^{7}$,
M. Clemens$^{9}$,
E. Corbelli$^{2}$,
I. De Looze$^{3}$,
S. di Serego Alighieri$^{2}$,
J. Fritz$^{3}$, \newauthor
G. Gavazzi$^{8}$,
C. Pappalardo$^{2}$, 
M. Grossi$^{6}$,
L. K. Hunt$^{2}$,
S. Madden$^{10}$,
L. Magrini$^{2}$, \newauthor
M. Pohlen$^{1}$,
M. W. L. Smith$^{1}$,
J. Verstappen$^{3}$ and
C. Vlahakis$^{11}$. \\ 
$^{1}$School of Physics and Astronomy, Cardiff University, The Parade, Cardiff, CF24
3AA, UK. \\
$^{2}$INAF-Osservatorio Astrofisico di Arcetri, Largo Enrico Fermi 5, 50125 Firenze, Italy. \\
$^{3}$Sterrenkundig Observatorium, Universiteit Gent, KrijgslAAn 281 S9, B-9000 Gent,
Belgium. \\
$^{4}$Jodrell Bank Centre for Astrophysics, School of Physics and Astronomy, 
University of Manchester, Oxford Road, Manchester M13 9PL, UK. \\
$^{5}$European Southern Observatory, Karl-Schwarzschild Str. 2, 85748 Garching bei Muenchen, Germany.  \\
$^{6}$
CAAUL, Observat\'orio Astron\'omico de Lisboa, Universidade de Lisboa,
Tapada da Ajuda, 1349-018, Lisboa, Portugal. \\
$^{7}$Laboratoire d'Astrophysique de Marseille, UMR 6110 CNRS, 38 rue F. Joliot-Curie,
F-13388 Marseille, France. \\
$^{8}$Universita' di Milano-Bicocca, piazza della Scienza 3, 20100, Milano, Italy. \\ 
$^{9}$INAF-Osservatorio Astronomico di Padova, Vicolo dell'Osservatorio 5, 35122 Padova,
Italy. \\
$^{10}$Laboratoire AIM, CEA/DSM- CNRS - Universit\'e Paris Diderot, Irfu/Service, Paris, France. \\
$^{11}$Departamento de Astronomia, Universidad de Chile, Casilla 36-D, Santiago,
Chile.
 }

\begin{document}

\date{Original January 2011}


\maketitle


\begin{abstract} We describe the Herschel Virgo Cluster Survey (HeViCS) and the first data that cover the complete survey area (four 4$\times$4 sq deg regions). We use these data to measure and compare the global far infrared properties of 78 optically bright galaxies that are selected at 500 $\mu$m and detected in all five far-infrared bands. We show that our measurements and calibration are broadly consistent with previous data obtained by IRAS, ISO, Spitzer and Planck. We use SPIRE and PACS photometry data to produce 100, 160, 250, 350 and 500 $\mu$m cluster luminosity distributions. These luminosity distributions are not power laws, but 'peaked', with small numbers of both faint and bright galaxies. We measure a cluster 100-500 $\mu$m far-infrared luminosity density of $1.6(7.0)\pm0.2\times10^{9}$ L$_{\odot}$ Mpc$^{-3}$.  This compares to a cluster 0.4-2.5 $\mu$m optical luminosity density of $5.0(20.0)\times10^{9}$ L$_{\odot}$ Mpc$^{-3}$, some 3.2(2.9) times larger than the far-infrared. A 'typical' photon originates from an optical depth of $0.4\pm0.1$. Most of our sample galaxies are well fitted by a single modified blackbody ($\beta=2$), leading to a mean dust mass of $\log{M_{Dust}}=7.31$ M$_{\odot}$ and temperature of 20.0K. We also derive both stellar and atomic hydrogen masses from which we calculate mean values for the stars:gas(atomic) and gas(atomic):dust mass ratios of 15.1 and 58.2 respectively. Using our derived dust, atomic gas and stellar masses we estimate cluster mass densities of $8.6(27.8) \times10^{6}$, $4.6(13.9) \times10^{8}$, $7.8(29.7) \times10^{9}$ M$_{\odot}$ Mpc$^{-3}$, respectively for dust, atomic gas and stars. These values are higher than those derived for field galaxies by factors of 39(126), 6(18) and 34(129) respectively. In the above luminosity/mass densities are given using the whole sample with values in brackets using just those galaxies that lie between 17 and 23 Mpc. We provide a data table of flux densities in all the Herschel bands for all 78 bright Virgo cluster galaxies.  \end{abstract}


\section{Introduction} 
The spectacular number of large diffuse nebulae in the constellation of Virgo has attracted the attention of astronomers for centuries. We now know the nebulae to be galaxies and that they reside in the largest nearby galaxy cluster. The cluster consists of many thousands of galaxies both physically large and small and because of its proximity ($v\sim$1094 km s$^{-1}$, Binggeli et al., 1987, $d\sim17$ Mpc, Gavazzi et al., 1999 and $d\sim16.5$ Mpc, Mei et al., 2007) the largest galaxies subtend some of the biggest angular sizes ($\sim$10 arc min) in the sky. 

These big bright galaxies occupy a relatively small area ($\approx 100$ sq deg) and so, over the years, the whole region has been subject to extensive 'survey' observations. Recent surveys range from the X-ray (Boehringer et al. 1994), Ultra-violet (Boselli et al. 2011), optical (VCC, Binggeli et al. 1985, VGVS, Mei et al. 2010, SDSS, Abazajian et al. 2009), near infrared (2MASS, Skrutskie et al. 2006, UKIDSS, Warren et al. 2007), far-infrared (IRAS, Neugebauer et al. 1984) and 21cm, (ALFALFA, Giovanelli et al. 2005, VIVA, Chug et al. 2009, AGES, Taylor 2010). These extensive large data sets provide us with a unique opportunity to study in detail a large number of galaxies at high spatial resolution. 

\begin{figure*}
\centering
\includegraphics[width=\textwidth]{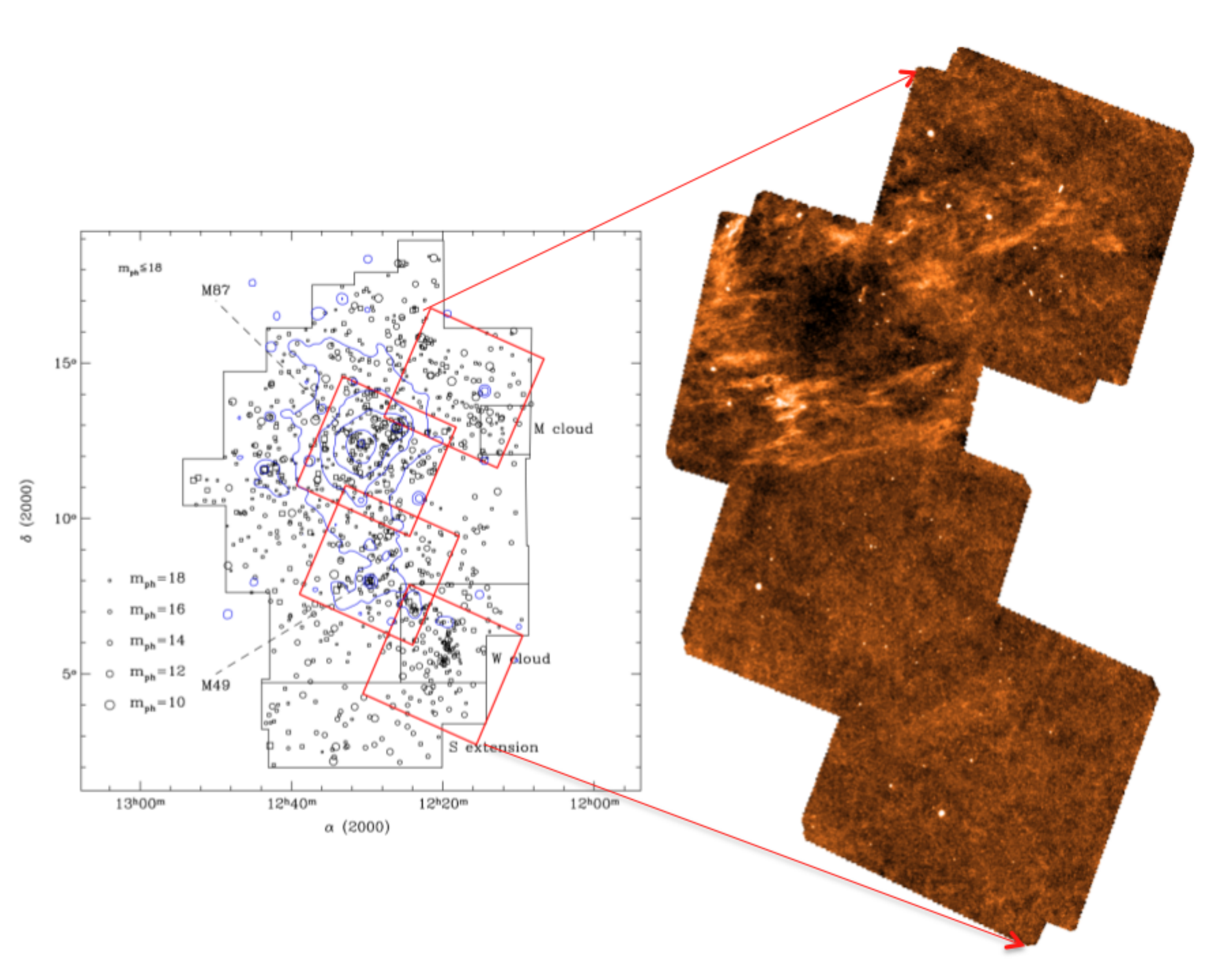}
\vspace{0.0cm}
\caption{The Virgo Cluster. On the left is shown the area ($\sim100$ sq deg) observed as part of the Binggeli et al. (1987) survey which led to the Virgo Cluster Catalogue (VCC) of some 2000 galaxies. Each VCC galaxy is marked as a circle. The area ($\sim64$ sq deg) covered by HeViCS is indicated by the red boxes. X-ray contours as observed by ROSAT are shown in blue (Boehringer et al., 1994). On the right are the four fields as observed by Herschel (SPIRE) at 250 $\mu$m. Bright galaxies can be seen as white dots. Prominent on the frames is the emission from Galactic cirrus.} 
\end{figure*}

However, the properties of cluster galaxies may not be typical of galaxies in general. The environment of a galaxy cluster is very different to that of other less densely populated 'field' regions of the Universe. The proximity of other galaxies and the existence of an inter-galactic medium means that Virgo cluster galaxies probably evolve in ways that are, in detail quite different from field galaxies. For example, in the cluster the morphological mix of galaxies is quite different from that in the field and many galaxies are observed to be relatively deficient in atomic gas (Haynes \& Giovanelli, 1984). This deficiency in gas is probably responsible for the observed lower star-formation rates (Lewis et al., 2002, Gomez et al., 2003) and the truncation of star-forming discs (Boselli \& Gavazzi, 2006) seen in cluster galaxies. An understanding of the prime physical processes that influence how a galaxy changes and evolves with time is a primary motivation for the study of galaxies in the nearby Universe, particularly the identification of processes unique to the cluster environment.

Given that the very essence of what a galaxy does is to form stars, and hence convert primordial gas into heavy elements (metals), the fractional mass of these elements compared to the gas is a primary indicator of how far a galaxy has proceeded along its evolutionary path. A major depository of these metals is in the interstellar dust and an important and outstanding legacy from the hugely successful IRAS survey of the 1980s has been the question of determining the total quantity of dust in galaxies. The issue has been outstanding because IRAS with a maximum wavelength of 100 $\mu$m, measured warm dust in galaxies (generally T$\sim30+$ K), leading to gas-to-dust mass ratios higher by about a factor of ten than that observed in our galaxy (measured by other means i.e. stellar extinction). They are also much higher than expected using chemical evolution models that predict the fractional mass of dust produced in each generation of stars (Edmunds and Eales 1998, Dwek 1998). The open question since then has been just how much cold dust (T$\le$20 K) remains undetected in galaxies and whether this can reconcile our local observations of the Milky Way and the dust production models.

There have of course been other far infrared space missions since IRAS that have to some extent addressed the cold dust issue. In particular, Tuffs et al. (2002) and Popescu et al. (2002) discuss this, amongst other issues, using bright Virgo cluster galaxies observed by the Infra-red Space Observatory (ISO). Other ISO observations of individual galaxies which specifically address the cold dust issue have been described by Alton et al., (1998a), Davies et al. (1999), Trewhella et al. (2000), Bendo et al. (2002) and Bendo et al. (2003). The Spitzer space telescope has also made observations of nearby galaxies at wavelengths at or just beyond that observed by IRAS (out to about 200 $\mu$m) with the Spitzer Infrared Nearby Galaxy Survey (SINGS) sample being of particular importance in helping us understand the far infrared properties of galaxies (Kennicutt et al., 2003). 

Both ISO and Spitzer have been somewhat limited in what can be inferred about the cold dust issue for three reasons. The first is that the wavelength coverage is not in the critical wavelength region where we expect to detect the signature of cold dust (200 - 600 $\mu$m). The second is that these were not survey telescopes and so they did not provide objectively selected large samples of galaxies for comparative study at the wavelengths of interest. Thirdly because small quantities of warm dust can swamp the signal from cold dust it is important to have sufficient spatial resolution to identify regions in the galaxy where cold dust emission is dominant. Although SCUBA observations at 850 $\mu$m (Dunne et al. 2000, Vlahakis et al. 2005, Alton et al. 1998b, Bianchi et al. 2000) have also attempted to address the above issues, with the advent of the Herschel Space Telescope these problems we hope can be solved conclusively. 

The observations of Virgo cluster galaxies described below make use of the unique imaging qualities and sensitivity of Herschel at five wavelengths from 100 - 500 $\mu$m. With a 3.5m mirror the spatial resolution of Herschel at these wavelengths ranges from 7 - 35 arc sec enabling spatially resolved observations of many of the bright galaxies at the distance of Virgo. The Herschel Virgo Cluster Survey (HeViCS) is an approved ESA Herschel Space Telescope (Pilbratt et al., 2010) Open Time Key Project. The project has
been awarded 286 hours of observing time in parallel mode using
PACS (Poglitsch et al., 2010) at  100 and 160 $\mu$m and SPIRE (Griffin et al., 2010)
at 250, 350 and 500 $\mu$m. Eventually we will map four 4$\times$4 sq deg regions of the cluster down to the confusion limit in the SPIRE bands. Pre-Herschel comparable surveys describing the far infrared properties of nearby bright galaxies have been described by Soifer et al. (1987), Doyon and Joseph (1989) (IRAS), Tuffs et al, (2002) (ISO), Draine et al. (2007) (Spitzer). 

In this paper we describe results from an initial data release that covers the entire HeViCS area but using only a quarter of the scans that will constitute the final data set. For this reason this paper concentrates on the properties of the bright galaxies because these are still detected at high signal-to-noise even in these as yet incomplete data. The primary HeViCS science goals using the full depth data include: the detection of dust in the inter-galactic medium, the extent of cold dust in the outskirts of galaxies, far-infrared luminosity functions, the complete spectral energy distributions (SEDs) of galaxies, the dust content of dwarf ellipticals and irregulars and a detailed analysis of the dust content of early type galaxies (for further details see http://www.hevics.org).

This paper is an extension of a previous paper (paper I, Davies et al., 2010) that considered the properties of the bright galaxies in a single central 4$\times$4 sq deg HeViCS field.  In a further seven papers on this central field we have discussed: how the cluster environment truncates the dust discs of spiral galaxies (paper II, Cortese et al., 2010), the dust life-time in early-type galaxies (paper III, Clemens et al., 2010), the spiral galaxy dust surface density and temperature distribution (paper IV, Smith et al., 2010), the properties of metal-poor star-forming dwarf galaxies (paper V, Grossi et al., 2010), the lack of thermal emission from the elliptical galaxy M87 (paper VI, Baes et al., 2010) and the far-infrared detection of dwarf elliptical galaxies (paper VII, De Looze et al., 2010). A further paper (Boselli et al., 2010) discusses the spectral energy distributions of HeViCS galaxies together with  others observed as part of the Herschel Reference Survey (HRS). 

\section{Observations, data reduction, object selection and calibration checks}
We have obtained $\sim64$ sq deg of data over four fields covering a large part of the 
Virgo Cluster using the SPIRE/PACS parallel scan-map mode (Fig 1). We use nominal
detector settings and a fast scan rate of 60\arcsec/sec over two orthogonal cross-linked
scan directions. Here we discuss combined data from two scans per field (the
  complete survey will comprise 8 scans).

PACS data reduction was carried out with the standard pipeline for both the 100
and 160 $\mu$m channels.
Dead and saturated pixels were masked. Deglitching was performed in two steps, using the standard multi-wavelength median transform 
deglitcher, and another based on
sigma-clipping. 
Bright sources were masked before a high-pass filter was used to reduce $1/f$ noise. 
Finally, the two orthogonal scans
were combined and maps made using the naive map-maker. Overall, the HeViCS PACS data
reduction strategy is similar to the approach used for the H-ATLAS key programme
(Eales et al. 2010), as explained in detail in Ibar et al. (2010).
After combining orthogonal scans the full width half maximum (FWHM) beam sizes are approximately 9\arcsec and 13\arcsec with pixel sizes of 3.2\arcsec and 6.4\arcsec for the 100
and 160 $\mu$m channels respectively. 

The SPIRE photometer (Griffin et al. 2010) data were processed up to Level-1
(to the level where the pointed photometer time-lines have been derived)
with a custom driven pipeline script adapted from the official pipeline
({\sl
POF5\_pipeline.py}, dated 8 Jun 2010) as provided by the SPIRE Instrument
Control Centre (ICC)
\footnote{See 'The SPIRE Analogue Signal Chain and
Photometer Detector Data Processing Pipeline' Griffin et al. 2009 or
Dowell et al. 2010 for a more detailed description of the pipeline and a list
of the individual modules.}.
This Jython script was run in the Herschel Interactive Processing
Environment (HIPE, Ott 2010). 
Our data reduction up to Level-1 is very
similar to
the Herschel Common Science System/Standard Product Generation v5 with a calibration based on Neptune data. 

Specific differences to the standard pipeline were that we used the {\it sigmaKappaDeglitcher} instead of the ICC-default 
{\it waveletDeglitcher}. Furthermore, we did not run the default {\it temperatureDriftCorrection} and the residual,
median baseline subtraction. Instead we use a custom method called BriGAdE (Smith et al. in preparation)
to remove the temperature drift and bring all bolometers to the same level (equivalent to baseline removal).
We have found this method improves the baseline subtraction significantly especially
in cases where there are
strong temperature variations during the observation.

Both scans were then combined to make our final maps using the naive mapper provided in the standard pipeline.
The FWHM of the SPIRE beams are
18.1\arcsec, 25.2\arcsec, and 36.9\arcsec with pixel sizes of 6\arcsec,
10\arcsec, and 14\arcsec at 250, 350, and 500 $\mu$m, respectively. 
The final data products have a mean 1$\sigma$ noise, determined from the whole of each image, of $\sim$2.6, 4.5, 0.9, 1.1 and 1.3 mJy pixel$^{-1}$ at 100, 160, 250, 350 and 500 $\mu$m respectively. 


To obtain a bright far-infrared selected sample we carried out our initial object selection at 500 $\mu$m.  This is because it is the least explored part of the spectrum, it has the lowest resolution, and most galaxies will produce their lowest flux in this band, guaranteeing a detection in all five bands. 
To produce an objectively selected sample we used the automatic image detection algorithm SExtractor (Bertin \& Arnouts, 1996). To minimise background contamination by faint sources each object was required to have more than 30 connected pixels at 1.5$\sigma_{500}$ or above. The result is a 500 $\mu$m flux density limit of $\sim$0.1 Jy for sources with a diameter larger than 1.4\arcmin. Each object was then checked for correspondence with a known Virgo Cluster galaxy. The final sample consists of 78 Virgo Cluster objects, which is 12\% of the 629 VCC galaxies in our fields that are listed by GOLDMINE as confirmed cluster members (Gavazzi et al., 2003). 
Discarded objects are either associated with extended galactic cirrus emission or objects in the background of Virgo. In this paper our intention is to concentrate on these bright galaxies detected at high signal-to-noise. The faintest galaxy in the final sample (see below) being detected with a signal-to-noise of $\sim$15.


Before carrying out aperture photometry on the selected galaxies the data 
at other wavelengths were smoothed and re-gridded to the 500 $\mu$m resolution 
and pixel scale. Elliptical apertures were initially chosen by eye using the 500 $\mu$m data, with the sky defined by a concentric annulus. These same annuli were then used on the smoothed and re-gridded data at other wavelengths.
Independent measurements were carried out on the maps at the original resolution,
defining for each galaxy a polygonal aperture that included all pixels with S/N$\ge$2
at 250 $\mu$m (the deepest band), and a nearby sky aperture that avoided pixels 
contaminated by bright background sources. Our results from these two approaches are significantly affected 
by the apertures chosen and after careful analysis we estimate errors of 17, 10, 8, 8 and 10\% at 100, 160, 
250, 350 and 500 $\mu$m respectively due to the choices made.

Independently as part of the HRS data analysis Ciesla et al. (in preparation) has compared our SPIRE aperture photometry with that derived using apertures set at 1.4 times the galaxy's optical radius and a 60\arcsec width annulus for the background. Mean values of the flux density ratios between these HRS values and those given here are $0.98\pm0.06$, $0.96\pm0.08$ and $0.96\pm0.09$ for 250, 350 and 500 $\mu$m respectively. The scatter between the Ciesla et al. values and those quoted here is 6, 8 and 10\% respectively, i.e. very close to the SPIRE aperture uncertainty we estimated above.

As can be seen from Fig. 1 the four separate $4\times4$ sq deg fields have a small overlap 
region. A few of our galaxies fall within these overlap regions and so have been 
observed independently twice. Also, two of the fields have been observed already
with more than two scans. Using the same apertures on these independent measurements 
leads to an uncertainty in flux values due to noise in the data recording and 
processing of 20, 10, 2, 3 and 5\% at 100, 160, 250, 350 and 500 $\mu$m respectively.


The absolute calibration of the data is thought to be uncertain to 15, 15, 7, 7 and 7\% at 100, 160, 250, 350 and 500 $\mu$m respectively (Ibar et al., 2010, Swinyard et al., 2010). Taking the above three uncertainties to be independent we estimate total uncertainties in our flux values of 30, 20, 10, 10 and 15\% at 100, 160, 250, 350 and 500 $\mu$m respectively. In table 1 we list all 78 Virgo bright galaxies with their names, co-ordinates, velocities, distances and flux values in each band. Co-ordinates are obtained from centroiding the 500 $\mu$m data, Velocities and distances have been taken from GOLDMINE (Gavazzi et al., 2003). The fluxes in Table 1 do not include color corrections (see Section 5).

\begin{figure}
\centering
\includegraphics[width=0.47\textwidth]{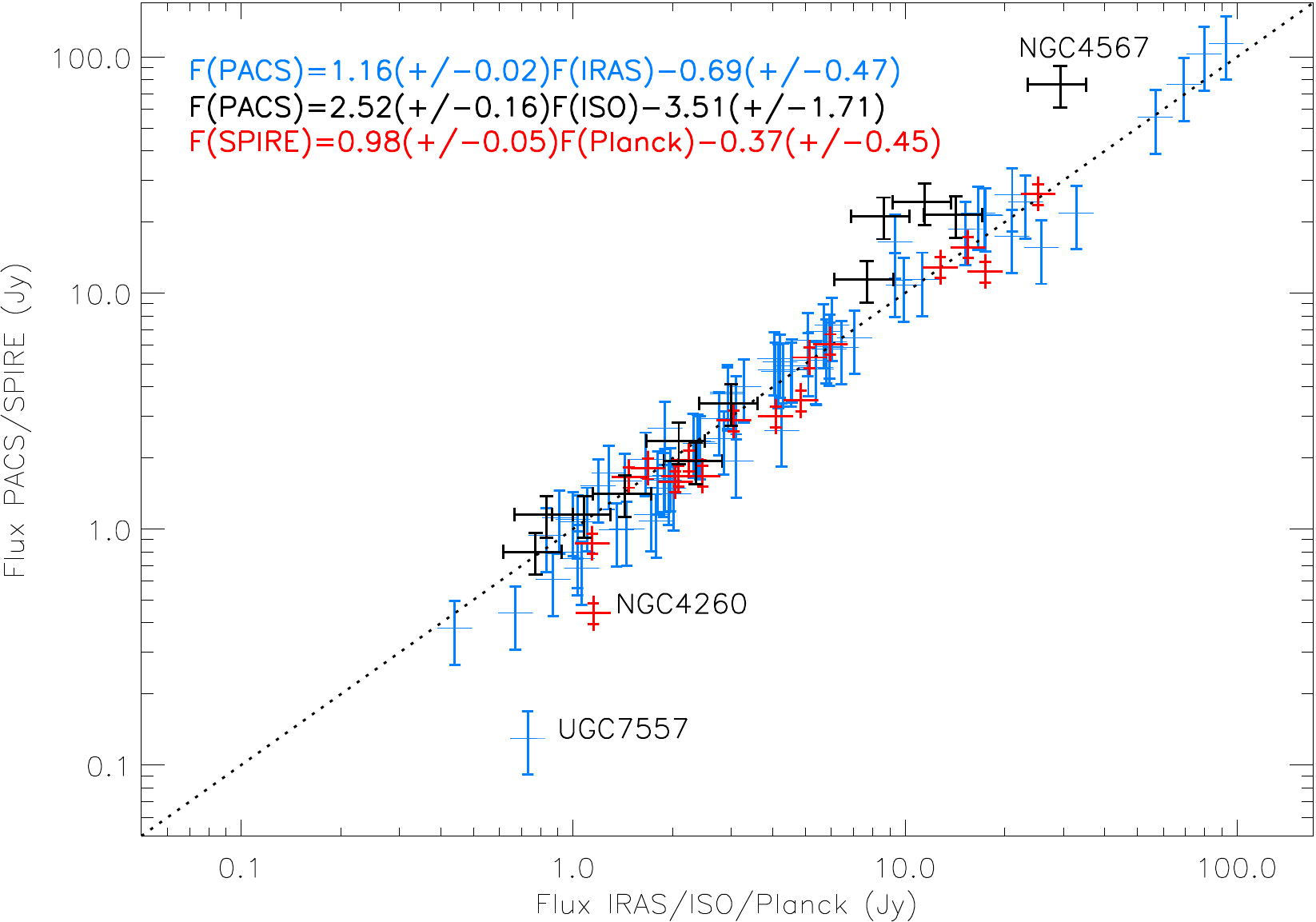}
\caption{A comparison of IRAS (blue) and ISO (black) fluxes with the PACS 100 $\mu$m fluxes along with the Planck/SPIRE 350 $\mu$m (red) fluxes. Linear least squares fit parameters are given top left. The black dashed line is the one to one relationship.
}
\end{figure}

For both PACS and SPIRE data we can make some comparisons of our results with those obtained by others. Where available (73 galaxies) we have taken IRAS 100 $\mu$m fluxes from the GOLDMINE database (Gavazzi et al., 2003) and show their comparison with PACS in Fig 2. UGC7557 stands out as being the only obvious anomalous result, as we measure a barely detected value of 0.13 Jy while the IRAS value is 0.73 Jy. A linear least squares fit to the data  gives the relationship $F$(PACS)=1.16($\pm0.02$)$F$(IRAS)-0.69($\pm0.47$). Using ISO Tuffs et al. (2002) also measured 100 $\mu$m fluxes for 12 of our galaxies, Fig. 2. The linear least squares fit to this data  gives the relationship $F$(PACS)=2.52($\pm0.16$)$F$(ISO)-3.51($\pm1.71$). This is the least good of our comparison calibrations with an indication that the largest fluxes correspond less well. NGC4567 has a PACS 100 $\mu$m flux a factor of 2.6 larger than that measured by ISO (76.3 compared with 29.3 Jy), Fig.2, but it is very close to NGC4568 and would be difficult to separate from it in the lower resolution ISO data. Spitzer 160 $\mu$m data exist for 44 of the 78 galaxies in our sample (Bendo, in preparation), Fig 3. A linear least squares fit to the data  gives $F$(PACS)=0.97($\pm0.02$)$F$(MIPS)+1.16($\pm0.56$). There appears to be just a small calibration off-set (PACS fluxes are higher) between PACS and MIPS. There are 14 galaxies from the ISOPHOT 170 $\mu$m Serendipity Survey (Stickel et al. 2004) that are in common with our survey, and we compare these fluxes with ours in Fig 3. Comparing these 14 with our 160 $\mu$m data we obtain $F$(PACS)=0.60($\pm0.05$)$F$(ISO)+2.4($\pm0.27$). Here there is a larger  discrepancy between the two data sets which appears to be entirely due to NGC4192 with the ISO measurement almost twice that of PACS (66.6 compared to 37.7 Jy). Removing NGC4192 from the sample gives $F$(PACS)=0.94($\pm0.05$)$F$(ISO)+0.16($\pm0.48$). There are also 12 galaxies measured at 170 $\mu$m in the ISO survey of Tuffs et al. (2002), Fig 3. A linear least squares fit to this data  gives $F$(PACS)=0.93($\pm0.05$)$F$(ISO)+2.11($\pm1.47$). Finally the Planck consortium have released a point source catalogue of bright sources which, at 350 $\mu$m, contains 17 of the galaxies in our list. In Fig 2. we compare these 350 $\mu$m flux densities (we use the flux within an aperture of radius=FWHM) with our SPIRE data obtaining the relationship $F$(SPIRE)=0.98($\pm0.05$)$F$(Planck)-0.37($\pm0.45$), only NGC4260 stands out as being an anomaly with regard to SPIRE and Planck.

\begin{figure}
\centering
\includegraphics[width=0.47\textwidth]{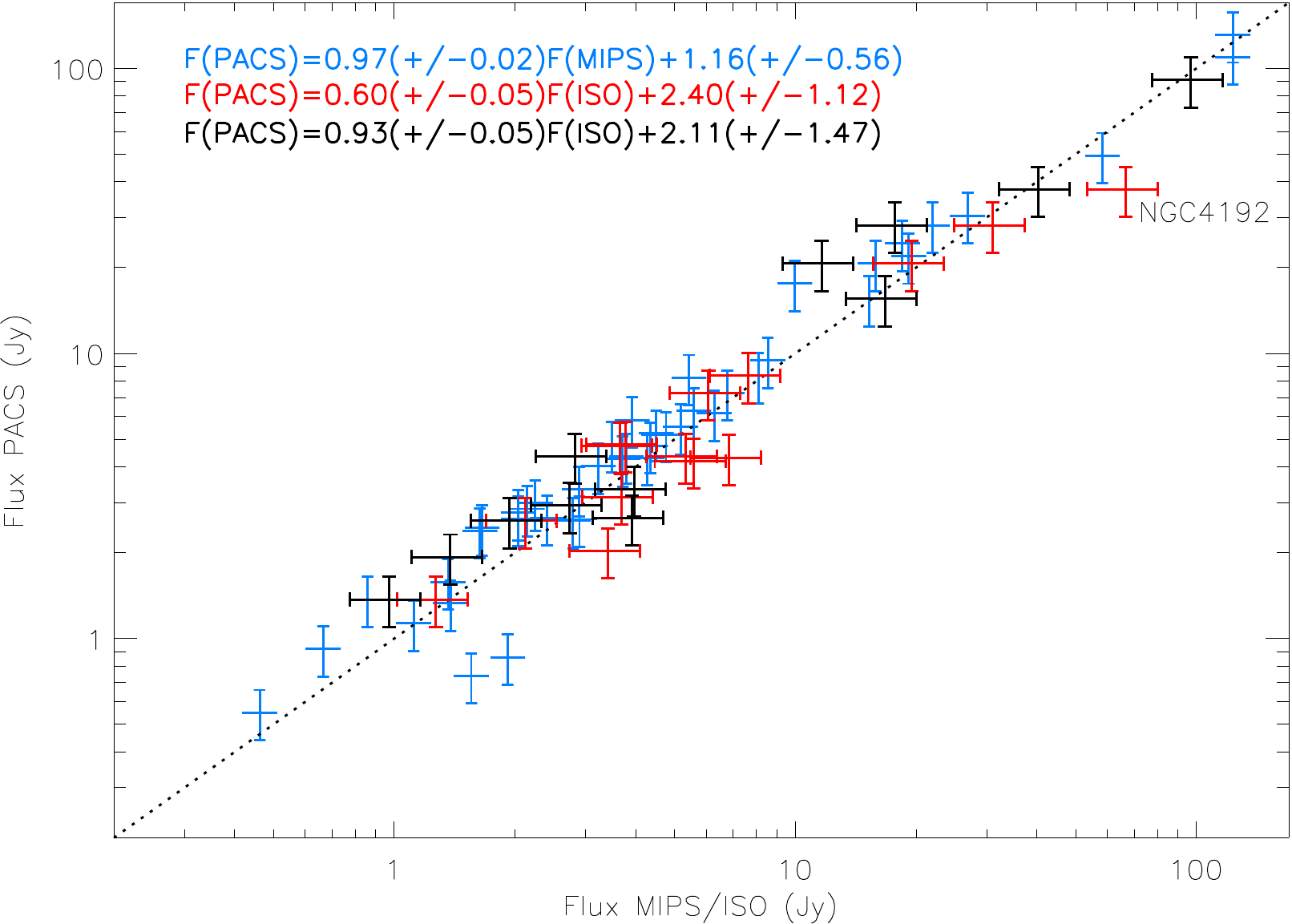}
\caption{A comparison of the Spitzer (MIPS) 160 $\mu$m (blue), the Stickel et al. ISO 170 $\mu$m (red) and the Tuffs et al, ISO 170 $\mu$m (black) fluxes with the PACS 160 $\mu$m fluxes. Linear least squares fit parameters are given top left. The black dashed line is the one to one relationship.
}
\end{figure}

\section{Luminosity functions and distributions}
Cosmic dust contributes only a small fraction of the total baryonic mass of galaxies, but  although small in mass, it plays a prominent role in how galaxies evolve with time. For example, dust is thought to be the major site for the formation of molecular hydrogen. Radiative cooling by dust grains allows this molecular gas to collapse to form stars and the residual dust is probably a major contributor to the material that forms proto-planetary discs and eventually planets. The link between dust and the formation of stars is confirmed by the close correlation of far-infrared emission at shorter wavelengths ($< 100$ $\mu$m) and the star formation rate measured by other means (e.g. Calzetti et al. 2010). Longer wavelength dust emission is probably associated with more extensive colder material distributed throughout galaxies (Bendo et al. 2010, Bendo et al. 2011). The comparative luminosity of a galaxy at different far-infrared wavelengths is thus a measure of the relative importance of the dust associated with star formation and that which is cold and diffusely distributed throughout the inter-stellar medium. 

The Virgo bright galaxy luminosity function/distributions (total numbers of galaxies within each luminosity interval) for each Herschel wavelength are shown in Fig. 4
\footnote{We use the term luminosity function to refer to the 500 $\mu$m data because the data is selected at this wavelength. We use the term luminosity distribution for data at other wavelengths.}. The shape of these luminosity distributions can be used in comparison with those obtained for galaxies in other environments to assess
the influence of that particular environment (field or varying richness of group or cluster) and
at different redshifts to assess evolution over cosmic time. Given the proximity of Virgo, and hence the detail and depth of the observations, we expect this bright galaxy sample to act as a bench mark for future studies. 

The far-infrared luminosity function/distributions we obtain are quite different from those derived in the optical. At optical wavelengths large numbers of low luminosity galaxies are found in the cluster that give rise to a reasonably good Schechter function fit with faint end slope of order -1.3 to -1.6 (Sabatini et al., 2003). All of our far-infrared luminosity function/distributions turn over at faint luminosities. 

All galaxy samples, no matter how selected, are subject to selection effects primarily related to the sensitivity of the observations compared to the surface brightness of the objects. Given that this is an isophotal size selected sample there may be galaxies that have large enough luminosities to appear in Fig. 4, but are not included in our sample because they are too small at the isophote of choice. Future studies will investigate this further so here we just discuss some relevant issues that indicate that our 500 $\mu$m luminosity function and derived luminosity distributions may not be too awry. 
\begin{enumerate}
\item Even the lowest luminosity galaxies in the sample are detected at a minimum signal-to-noise of 15 i.e. the last point of the 500 $\mu$m luminosity function is not adversely affected by being close to a limit set by signal-to-noise.
\item In Davies et al. (2010) we show that the luminosity function/distributions are peaked independent of the wavelength (for example 160 $\mu$m) used for selection. 
\item At optical wavelengths Sandage et al., (1985) found that the luminosity distributions of Virgo cluster bright spiral and elliptical galaxies (analogous to our sample) were Gaussian and only became power laws (Schechter) when the dwarf galaxies were included. These dwarf galaxies are predominately metal poor dE galaxies and we have previously shown that these are not detected at 500 $\mu$m in our current data (De Looze et al., 2010). 
\item Confirmed Virgo cluster galaxies fainter than our 500 $\mu$m selection limit have been detected in the HeViCS data, but they are too small to be included in our sample (Grossi et al., 2010). Grossi (private communication) has identified 24 out of 139 late type dwarf galaxies listed in the VCC that have a measurable 500 $\mu$m flux. Six of these lie, according to GOLDMINE, in the most distant structure at 32 Mpc. We have added these 24 galaxies to the 500 $\mu$m data shown in Fig. 4 (yellow dashed line) and as can be seen they do not alter our conclusion about a 'peaked' luminosity function.
\end{enumerate}
 We have an on-going programme to try to detect as many as possible VCC galaxies in the HeViCs data. For this programme to seriously change our conclusions we would need to find a population of far-infrared bright objects that have very low optical luminosities. There is of course always the possibility of far-infrared sources with no optical counterparts, but we would have to show conclusively that they were cluster members and not associated with faint optical sources in the background.

If confirmed the 'peaked' nature of the far-infrared luminosity function/distributions is similar to the Virgo cluster HI mass function, where there is also an apparent turnover at low masses (Davies et al., 2004, Taylor et al., 2010). This is also quite different to the low mass power law slope of the global HI mass function (Martin et al. 2010, Davies et al. 2011). The straight forward explanation of the different HI mass function shapes is that gas stripping processes (Doyon \& Joseph, 1989) in the cluster are more dramatic for low mass galaxies. So the form of the luminosity function/distributions suggests that this may now also apply to the removal of cosmic dust. Although a number of tidal gas (21cm) streams (Kent et al., 2007) have been identified in the cluster environment the identification of inter-galactic dust has been a little more controversial and would benefit from additional confirmation (see Stickler et al. 1998 for an example of a detection). The fate of stripped galactic dust is still unclear. 

\begin{figure*}
\centering
\includegraphics[width=0.75\textwidth]{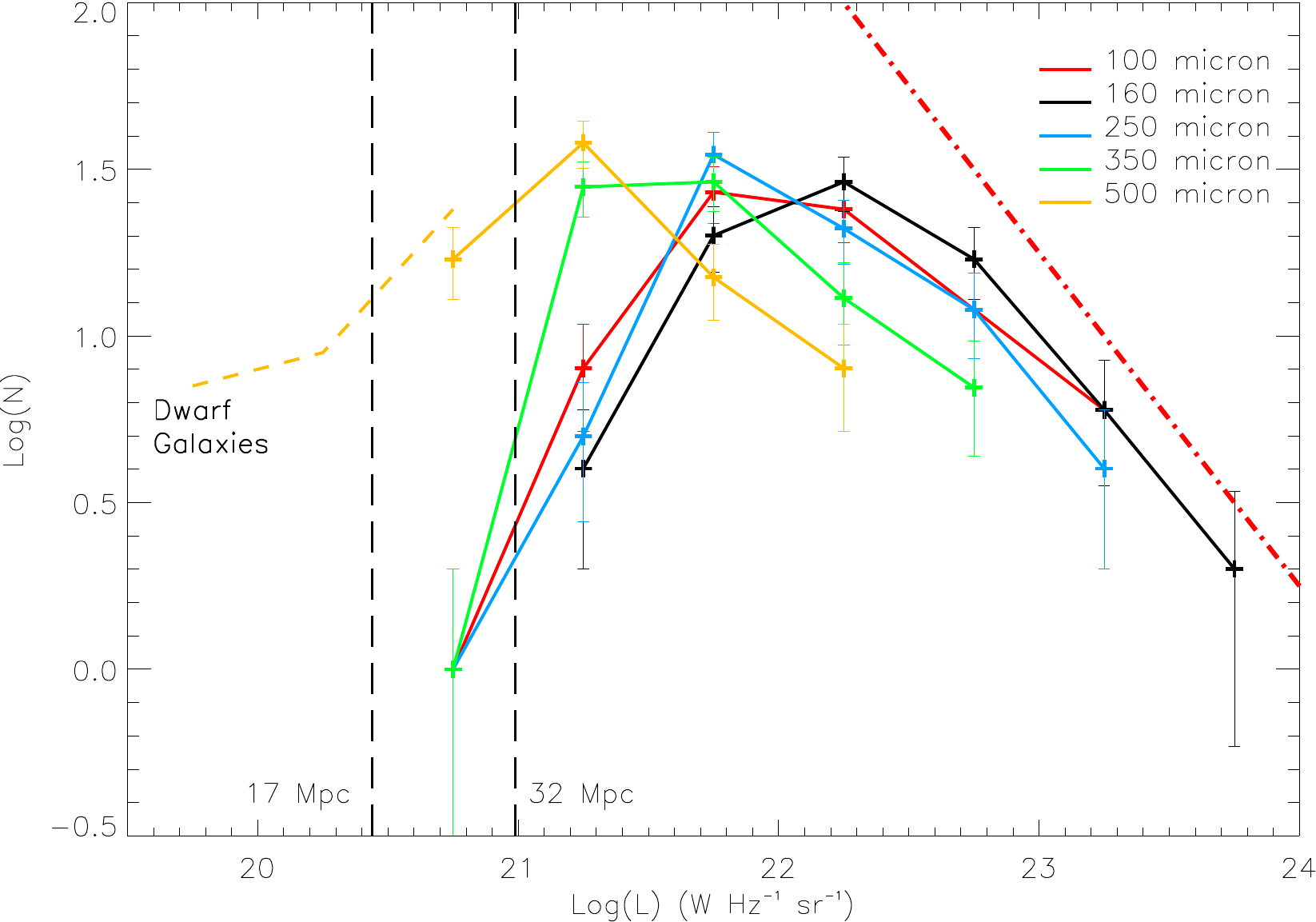}
\caption{The luminosity distributions derived from the data listed in table 1. The dot-dashed line indicates a slope of -1 as found by Rowan-Robinson et al. (1987) using IRAS 100 $\mu$m data. The vertical black dashed lines indicate the minimum luminosity at 17 and 32 Mpc for a minimum 500 $\mu$m flux density of 0.1 Jy. The yellow dashed line indicates the shape of the 500 $\mu$m luminosity distribution if known star forming dwarf galaxies, beyond our detection limit, are included. } 
\end{figure*}

At the current time the availability of comparison luminosity distributions derived at the wavelengths of interest here is somewhat limited (see Davies et al., 2010). Many of the luminosity distributions derived using IRAS data used the 60 $\mu$m fluxes as their longest wavelength i.e. Soifer et al. (1987), Saunders et al. (1990). With a few exceptions (see Takeuchi et al. (2006) and references therein) ISO also concentrated on shorter wavelengths than those considered here. The same is true of Spitzer (Babbedge et al. 2006, Rodighiero et al. 2010)

The most informative comparison is with the IRAS 100 $\mu$m luminosity function derived by Rowan-Robinson et al. (1987). In a similar way to us they constructed a far-infrared sample from optically identified galaxies with an available redshift. This is roughly comparable with our Virgo sample in the sense that the galaxies all have to have an optical identification and a distance, but their sample is drawn from a wide range of environments. The Rowan-Robinson et al. luminosity function follows a power law slope of about -1 from $\sim 10^{24}$ to $\sim 4\times10^{21}$ W Hz$^{-1}$ sr$^{-1}$.  Although the luminosity function/distribution normalisation is arbitrary when comparing to our data (Fig. 4) it is clear that the Rowan-Robinson et al. luminosity function predicts $\sim10^{3}$ more galaxies at the faint end than at the bright end - even with the small numbers in our sample this is difficult to reconcile with our Virgo data, unless subsequent investigation reveals large numbers of small faint cluster far infrared sources.  The Rowan-Robinson et al. data also include galaxies that are one order of magnitude brighter than the brightest galaxies found in the Virgo cluster - it has previously been noted that clusters in general do not have the very bright far-infrared sources seen in the field (Bicay \& Giovanelli, 1987).

Given the apparent shape of the Virgo cluster luminosity function/distributions it is currently more informative to simply characterise them by their mean values rather than a poorly fitted and parameterised Schechter function. These mean values are given in table 2. It is clear that the majority of the far-infrared energy of the cluster is being produced at the shortest of these wavelengths. Using Wien's law for the predicted temperature of a modified blackbody curve with emissitvity $\beta$, i.e. $T_{d}\approx \frac{14142.9}{(5+\beta)\lambda_{max}}$, leads to the temperatures given in table 2 for each wavelength (band) $\lambda_{max}$ and emissivity $\beta=2$. Clearly we expect from this dust temperatures of order 13-20K (see below). 

As the luminosity distributions apparently turn over at both the faint and luminous ends we can make an estimate of the luminosity density in each band. Listed Virgo members in GOLDMINE extend from 17 to 32 Mpc, which is a large distance compared to what is normally assumed for the size of a cluster. So, all densities quoted below will be calculated using both the whole sample and just using those with distances between 23 and 17 Mpc (given in brackets). The volume we sample over 64 sq deg of sky is $\sim$181.3 Mpc$^{3}$ for galaxies between 17 and 32 Mpc and $\sim$47.2 Mpc$^{-3}$ for the reduced sample with distance between 17 and 23 Mpc (68 galaxies). Values for the luminosity density in each band are given in table 2. It is difficult to make comparisons with previous work because this is a newly explored part of the electro-magnetic spectrum. Saunders et al., (1990) using IRAS data give a luminosity density of $4.0\pm0.4\times10^{7}$ L$_{\odot}$ Mpc$^{-3}$ for emission in the range 42.5-122.5 $\mu$m from galaxies in all environments. Using this band pass and our 100 $\mu$m luminosity density gives a value of $1.1(3.3)\pm0.1\times10^{9}$ L$_{\odot}$ Mpc$^{-3}$  for this Virgo cluster sample, a factor of 28(83) higher. At 160 $\mu$m Takeuchi et al. (2006) use ISO data to obtain a value of $3.7\times10^{7}$ L$_{\odot}$ Mpc$^{-3}$ again for a sample occupying various environments. Our value for Virgo is $5.1(15.9)\pm0.6\times10^{8}$ L$_{\odot}$ Mpc$^{-3}$, a factor of 14(44) higher. A value of H$_{0}$=72 km s$^{-1}$ Mpc$^{-1}$ has been assumed to adjust the literature luminosity densities. 

Simply multiplying the luminosity densities given in table 2 by the band widths (assuming non-overlapping bands, see below) we obtain a Virgo 100-500 $\mu$m far-infrared luminosity density of $\rho_{100-500}=2.2(9.6)\pm0.3 \times10^{-32}$ W m$^{-3}$ or $1.6(7.0)\pm0.2\times10^{9}$ L$_{\odot}$ Mpc$^{-3}$ using a solar luminosity of $3.9\times10^{26}$ W.
\setcounter{table}{1}
\begin{table}
\begin{center}
\begin{tabular}{c|ccc}
Band    &  Mean luminosity &  Luminosity density & Temperature \\
   ($\mu$m)  & $\times 10^{22}$ (W Hz$^{-1}$ sr$^{-1}$) & $\times 10^{-45}$ (W m$^{-3}$ Hz$^{-1}$) & (K) \\ \hline
100 & $3.4\pm0.6$  & 6.5(19.7)$\pm0.6$ & 20.2 \\
160 & $4.4\pm0.8$  & 8.3(25.9)$\pm0.6$ & 12.6 \\
250 & $2.4\pm0.4$  & 4.6(14.8)$\pm0.2$ & 8.1 \\
350 & $1.0\pm0.2$  & 1.9(6.3)$\pm0.2$ & 5.8  \\
500 & $0.4\pm0.1$  & 0.7(2.3)$\pm0.2$ & 4.0 \\
\end{tabular}
\caption{The mean luminosity and luminosity density in each band for the Virgo bright galaxy sample. The luminosity density using just the galaxies with distances of 17-23 Mpc are given in brackets. The temperatures are those that produce modified blackbody curves that peak in each band ($\beta=2$).}
\end{center}
\end{table}

\section{Far-infrared and optical luminosity}
The luminosity density ($\rho_{100-500}$) given above was calculated using non-overlapping bands corresponding to each Herschel wavelength (not necessarily centred on the nominal wavelength because of the uneven band spacing). This also provides a means of calculating a far-infrared luminosity for each galaxy. We define the Herschel far-infrared luminosity as: \\
\begin{center}
$L_{100-500}=3.1\times10^{4}d_{Mpc}^{2}[(f_{100}\Delta f_{100})+(f_{160}\Delta f_{160})+(f_{250}\Delta f_{250})$ \\
$+(f_{350}\Delta f_{350})+(f_{500}\Delta f_{500})]$ \hspace{0.5cm} $L_{\odot}$
\end{center}
where $f_{100}$, $f_{160}$, $f_{250}$, $f_{350}$, $f_{500}$, $\Delta f_{100}=18.0$, $\Delta f_{160}=8.9$, $\Delta f_{250}=4.6$, $\Delta f_{350}=3.1$ and $\Delta f_{500}=1.8$ are the flux density (Jy) and bandwidth ($10^{11}$ Hz) in each band respectively and we have again taken the solar luminosity to be $3.9\times10^{26}$ W. For this particular analysis we prefer this method of calculating the luminosity because it does not depend on any particular fit to the SED i.e. it is valid for galaxies well fitted by a single modified blackbody with any value of emissivity power index ($\beta$), those that require two components or more and those that do not have a thermal spectrum. 

To check the consistency of our determination of the far-infrared luminosity we have compared it with that predicted analytically for a galaxy that is well fitted by a modified blackbody (see next section). For NGC4254 the analytical value derived from the fit gives $\log{(L_{FIR})}=10.59$ while the above method gives $\log{(L_{100-500})}=10.51$.  Far-infrared luminosities ($L_{100-500}$) for each galaxy are listed in table 3. Comparing the average far-infrared SED of the galaxies in the sample, we find the highest flux density (38\%) in the 160 $\mu$m band.

We can use a similar approach for the apparent stellar luminosity of each galaxy. We define the apparent stellar luminosity as: \\
\begin{center}
$L_{0.4-2.5}=3.1\times10^{7}d_{Mpc}^{2}[(f_{g}\Delta f_{g})+(f_{r}\Delta f_{r})+(f_{i}\Delta f_{i})$ \\
$+(f_{J}\Delta f_{J})+(f_{H}\Delta f_{H}+(f_{K}\Delta f_{K}))]$ \hspace{0.5cm} $L_{\odot}$
\end{center}
where $f_{g}$, $f_{r}$, $f_{i}$, $f_{J}$, $f_{H}$, $f_{K}$, $\Delta f_{g}=1.9$, $\Delta f_{r}=1.1$, $\Delta f_{i}=1.6$, $\Delta f_{J}=0.9$, $\Delta f_{H}=0.3$ and $\Delta f_{K}=0.5$ are the flux (Jy) and bandwidth ($10^{14}$ Hz) in each band respectively. The $g$, $r$ and $i$ band data ($\lambda_{g}=0.48\mu m$, $\lambda_{r}=0.62\mu m$, $\lambda_{i}=0.76\mu m$) are taken from the Sloan Digital Sky Survey (SDSS). Automated magnitude derivations from within the SDSS database are notoriously inaccurate for extended sources, so for 54 sources where the data is available we have used SDSS magnitudes derived from our own surface photometry as described in Cortese et al. (2011). For the additional 24 galaxies we have used SDSS data from the NASA Extra-galactic Database (NED). For most of these galaxies, the magnitudes are plausibly consistent with those given in the near-infrared. For two galaxies (UGC7557 and NGC4451) there are no consistent SDSS magnitudes and we have set their apparent luminosities in these bands to zero. As we will see below this does not greatly affect our calculations because by far the largest fraction of the apparent stellar luminosity is emitted in the near-infrared. The SDSS AB magnitudes are converted to Jy using a zero point of $K_{SDSS}=8.9$. In the near infrared we have used the two Micron All Sky Survey (2MASS) which lists J, H and K band (total) magnitudes for all of our galaxies ($\lambda_{J}=1.25\mu m$, $\lambda_{H}=1.65\mu m$, $\lambda_{K}=2.17\mu m$). The 2MASS website gives the following zero points for the conversion of magnitudes to Jy: $K_{J}=8.01$, $K_{H}=7.53$, $K_{K}=7.06$.

Considering the average SED of all of the galaxies in the sample, but this time in the optical, the highest flux density is produced in the  $H$ band (28\% of the total), 23 and 22\% are produced in the $J$ and $K$ bands respectively, thus justifying our statement above that by far the largest fraction of the escaping star light is emitted in the near-infrared bands.

Summing the contribution from all the galaxies leads to an optical luminosity density of $\rho_{0.4-2.5} = 5.0(20.0)\times10^{9}$ L$_{\odot}$ Mpc$^{-3}$ some 3.1(2.9) times larger than the far-infrared value given in the previous section. For their ISO sample of Virgo galaxies Popescu et al. (2002) obtain values of 5.7 and 2.3 for the ratio of optical to far infrared luminosity in early and late type galaxies respectively, consistent with our global value. There are of course some very optically luminous galaxies like NGC4486 (M87) and NGC4374 (M84)  that emit little in the far-infrared. The range of both optical and far-infrared luminosities for the galaxies in this sample is shown in Fig. 5.

We can use the optical and far-infrared luminosities to make a crude estimate of the 'typical' optical depth ($<\tau>$) experienced by a photon as it leaves a galaxy, based on a simple screen of dust model:
\begin{center}
$<\tau>=\ln{\left(1.0+\frac{L_{100-500}}{L_{0.4-2.5}}\right)}$ \\
\end{center}
Values of $<\tau>$ are also listed in table 4. The mean value for galaxies in this sample is $<\tau>_{mean}=0.4\pm0.1$ so on average the optical energy is emerging from regions of intermediate optical depth (neither totally optically thin or thick). This mean value may be high when compared to galaxies in general because this is a far infrared selected sample. Values of $<\tau>$ range from 0.04 to 2.76. The two 'optically thick' galaxies with $<\tau>$ greater than unity are both late type spirals - NGC4234 (Sc) and NGC4299 (Scd). As galactic dust is typically confined to a relatively thin disc, the value of $\tau$ should be dependent on the inclination of each galaxy to the line of sight. It is therefore surprising that both NGC4234 and NGC4299 are relatively face-on galaxies with inclinations of approximately 40 and 20 degrees respectively. In the optical both galaxies show signs of possible disturbance and so maybe the dust has been 'stirred up' in these two galaxies. Given the complexities of the relative distributions of stars and dust a full understanding of the radiative processes that give rise to dust emission will only come about by carrying out detailed radiative transfer modelling (e.g. Bianchi et al., 2000a, Bianchi, 2008 and Baes and Dejonghe, 2001). The above value of $<\tau>_{mean}$ is very close to the value given in Saunders et al. (1990) derived in a similar way using IRAS data ($<\tau>_{mean}^{IRAS} = 0.3\pm0.1$).
The value we find implies than on average $33\pm7$ \% of the stellar radiation of a galaxy is absorbed by dust. The value is in
agreement with what was found by Popescu and Tuffs (2002) using a sample of late-type Virgo galaxies observed with ISO.
After removing a contribution of hotter dust using data at $60\mu$m, they fitted the 100 and 170 $\mu$m fluxes
with a single temperature modified blackbody with $\beta=2$ and extrapolated to longer wavelength. Using
the Herschel data up to 500 $\mu$m, we will show in the next Section that indeed this simple model can describe 
very well the dusty thermal SED in most of our sample.

\begin{figure}
\centering
\includegraphics[width=0.47\textwidth]{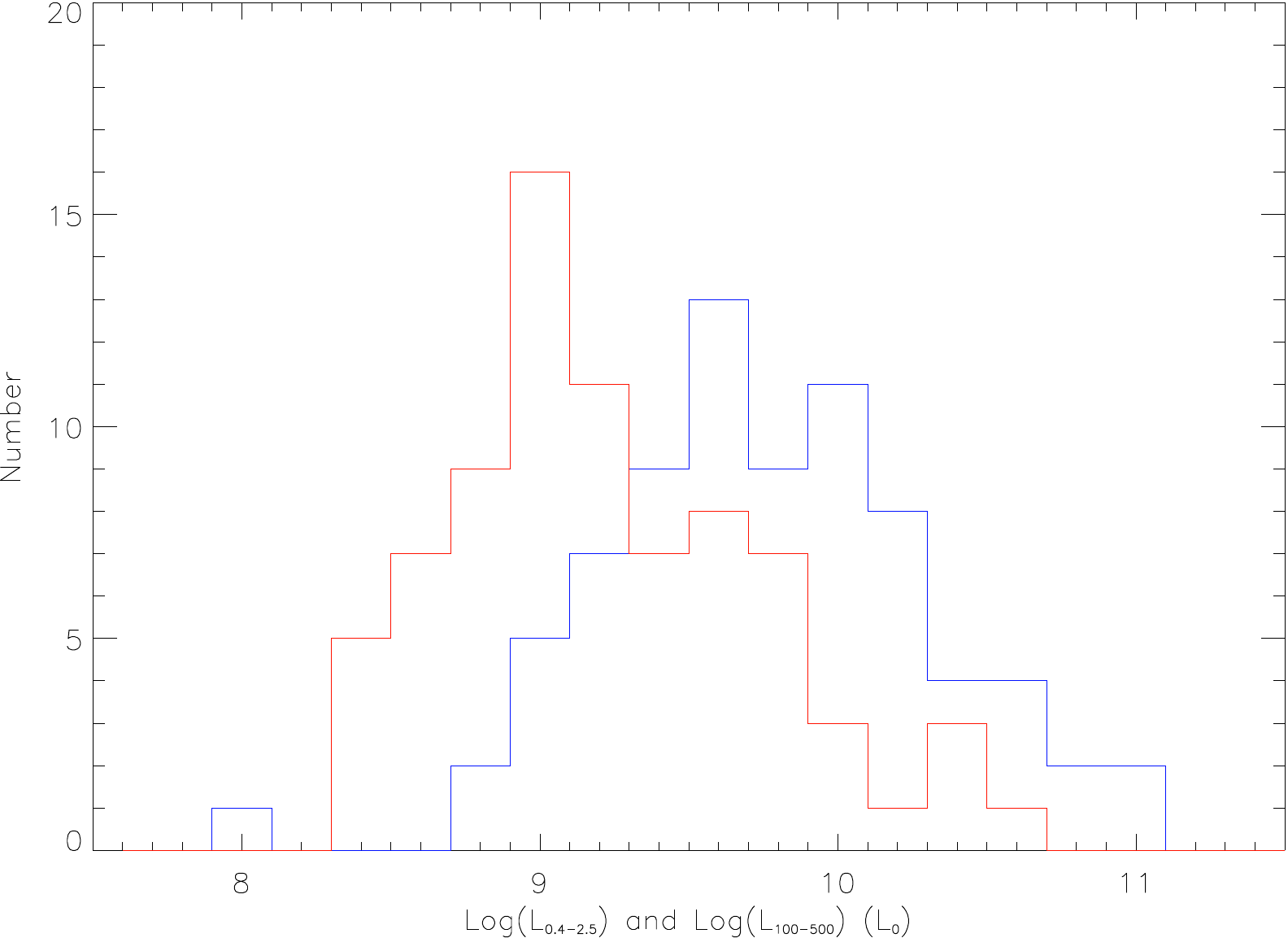}
\caption{A comparison of the range of optical luminosity in solar units from 0.4 - 2.5 $\mu$m (blue) with that in the far-infrared from 100 - 500 $\mu$m (red).}
\end{figure}

\section{Dust, gas and stellar mass}

\begin{figure*}
\centering
\includegraphics[width=16cm]{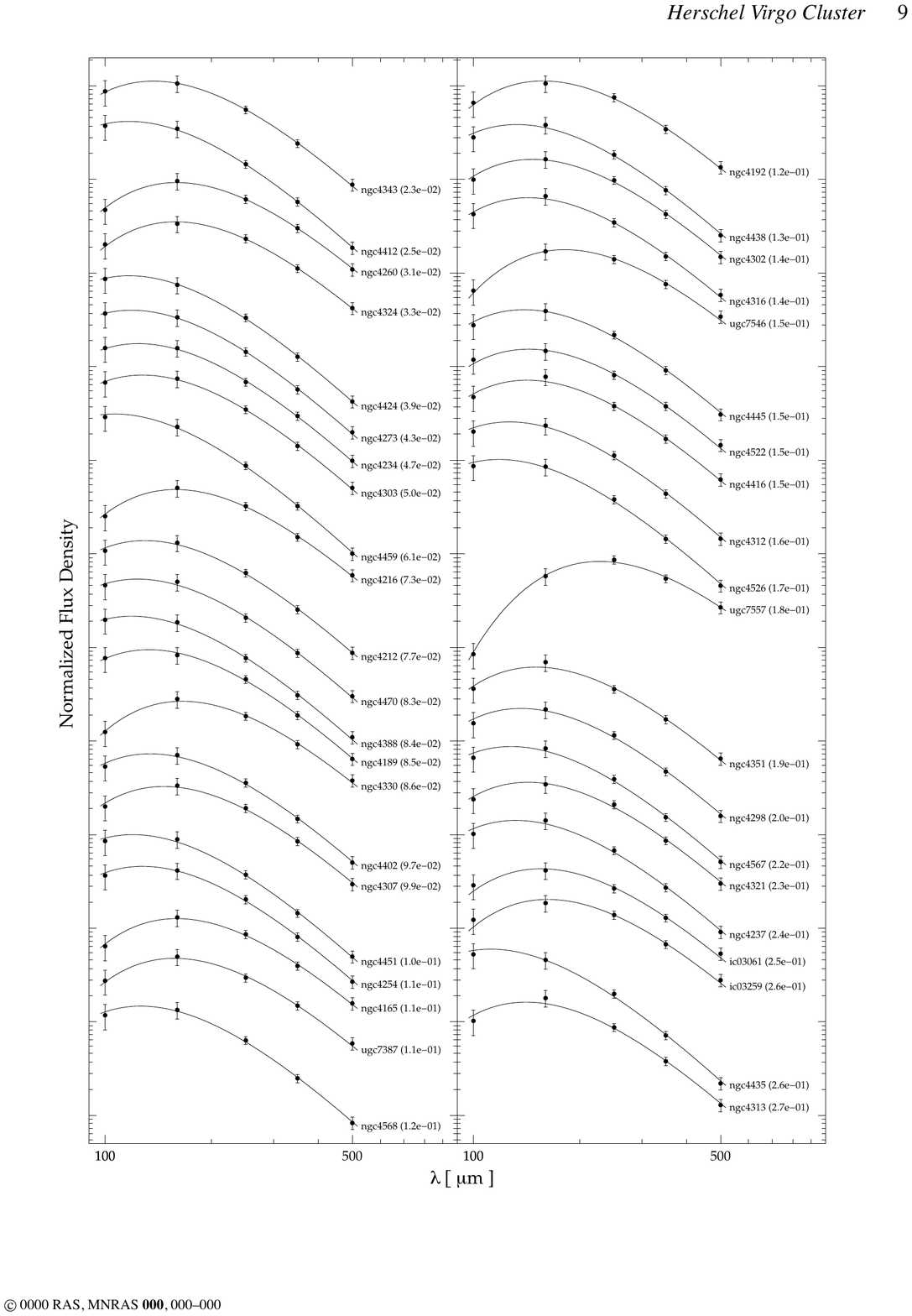}
\end{figure*}
\begin{figure*}
\centering
\includegraphics[width=16cm]{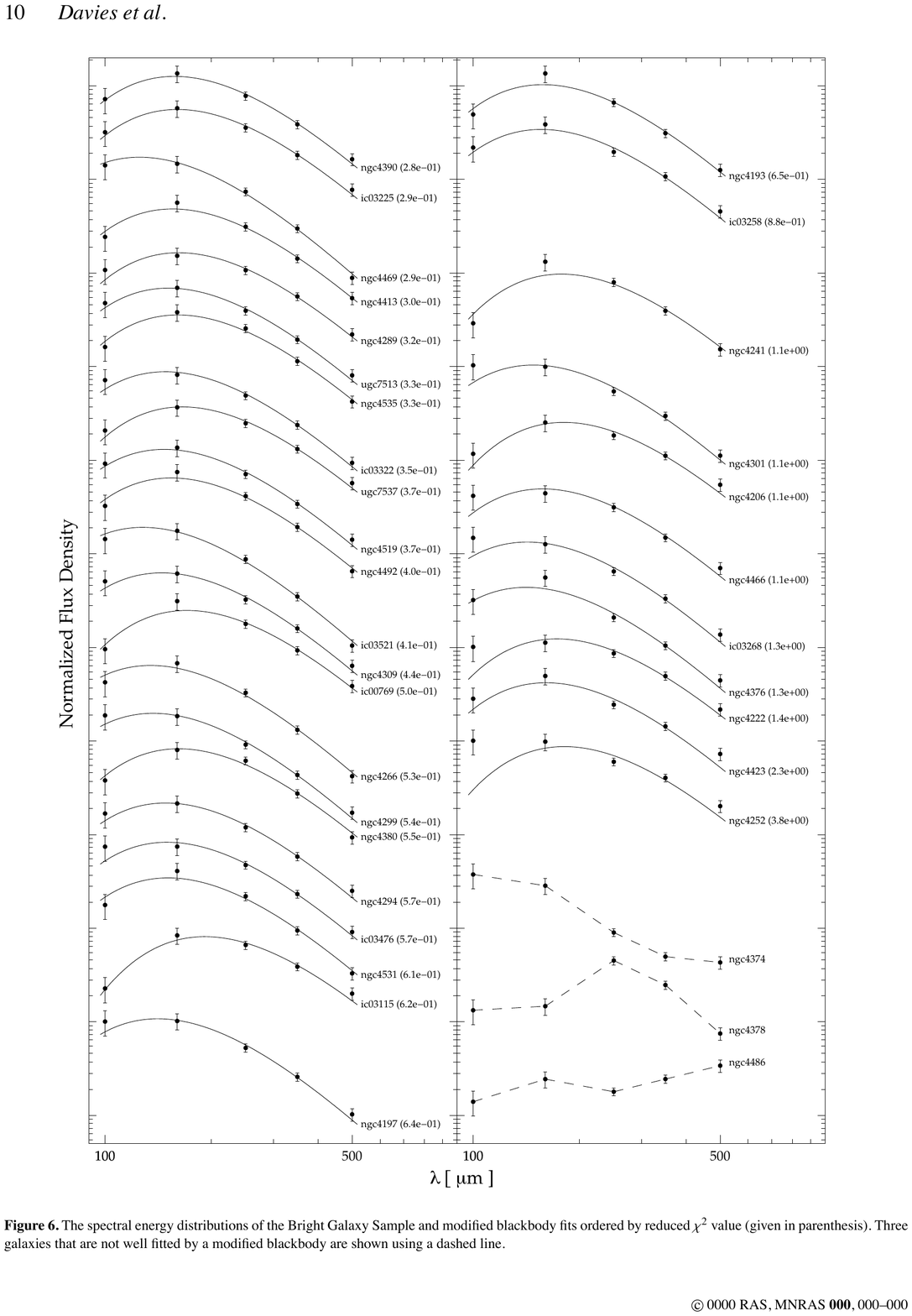}
\caption{The spectral energy distributions of the Bright Galaxy Sample and modified blackbody fits ordered by reduced $\chi^{2}$ value (given in parenthesis). Three galaxies that are not well fitted by a modified blackbody are shown using a dashed line.}
\end{figure*}


In order to derive the dust temperature and masses, we fitted the
SED for each galaxy, as defined by the Herschel flux densities in Table 1,
with a single temperature modified blackbody. We followed the
same procedure adopted in Smith et al. (2010), Bendo et al. (2010)
and Magrini et al. (2011), (see this later work for a discussion of the
uncertainties when using a single temperature model). 

We adopted a power 
law dust emissivity $\kappa_\lambda=\kappa_0 (\lambda_0/\lambda)^\beta$,
with spectral index $\beta=2$ and emissivity $\kappa_0$ = 0.192 $m^2$ kg$^{-1}$
at $\lambda_0$ = 350 $\mu$m. These values reproduce the behaviour of models
of Milky Way dust in the FIR-submm, which were also found to reproduce
the dust emission SED in several other galaxies (Draine 2003, 
Draine et al. 2007). The fit was obtained with a standard $\chi^2$ 
minimization technique. The monocromatic flux densities measured from
the images, and reported in Table 1, come from the pipeline
calibration. They have been derived from the passband-weighted flux 
density (measured by the instruments), applying a color correction
for a flat energy spectrum ($F_\nu \propto \nu^{-1} $). We retrieved the
passband-weighted flux density by removing this correction (for SPIRE, this 
is equivalent to dividing the pipeline flux densities by the $K_4$ factor 
for point sources; SPIRE Observer' Manual, 2010) and then fitted to the
data the average of the model over the spectral response function for each 
of the bands. We used the appropriate response functions for the PACS and 
SPIRE bands.
The above procedure is equivalent to applying color corrections to the
pipeline monochromatic fluxes, if a SED model was known a priori.
The color corrections are given in Table~\ref{tabcc}. Given the nature of
most of our sources, we used the SPIRE spectral response functions for
extended emission. Since these corrections are much smaller than the assumed 
errors in each band, the fluxes given in Table 1 have not been color 
corrected. In a few of our galaxies, the SPIRE fluxes might be dominated
by point source emission and thus need a larger color correction (see
Table 3). Even in this case, the color corrections
are of order or less than the errors derived above.

\begin{table} 
\begin{center}
\label{tabcc}
\begin{tabular}{@{}lccccc} 
\hline 
T (K) & 100 $\mu$m &160 $\mu$m &250 $\mu$m &350 $\mu$m &500 $\mu$m \\
\hline 
15 & 0.96 & 1.06 & 1.00   & 0.99   & 0.99 \\
   &      &      & (0.98) & (0.96) & (0.92) \\
20 & 1.03 & 1.03 & 0.99 & 0.98 & 0.98 \\
   &      &      & (0.96) & (0.94) & (0.90) \\
25 & 1.03 & 0.99 & 0.08 & 0.97 & 0.97 \\
   &      &      & (0.94) & (0.93) & (0.88) \\
\hline
\end{tabular}
\caption{Colour corrections for a $\beta=2$ modified blackbody of temperature T. Pipeline fluxes (i.e. calibrated assuming $\nu F_\nu = \mathrm{const.}$)
should be multiplied by the above factors to retrieve the corrected fluxes.
For the SPIRE bands, the value outside brackets are for extended sources,
while those within brackets are for point sources.} 
\end{center}
\end{table}

Remarkably, plotting the data given in table 1 shows that the majority of galaxies are simply and well fitted by this single temperature (modified) blackbody curve with fixed emissivity index; the fits are illustrated in Fig. 6 (ordered by $\chi^{2}$ value). There are three notable exceptions to this. The early type galaxies NGC4486 (M87) and NGC4374 (M84) show departures from a thermal spectrum (Boselli et al. 2010) particularly at longer wavelengths. As discussed in Baes et al., (2010) specifically for NGC4486, both these galaxies require modelling of the synchrotron component and its removal before the thermal dust spectrum can be analysed. 
The third galaxy, NGC4378, is an early type spiral with a Seyfert nucleus. Its SED does not appear to be contaminated by a non-thermal contribution, but rather dominated by a colder dust component, because its 160 $\mu$m flux is lower than expected when compared to the other values. NGC4378 requires more detailed radiative transfer modelling and probably a multi-component dust model before a substantial cold dust component is confirmed.
Further modelling of these galaxies with peculiar SEDs is deferred to future papers. Here we concentrate on dust emission and those galaxies that can be fitted by a thermal spectrum. 

A few galaxies do not fit a thermal spectrum as well as others, though they 
do not show any clear evidence of a non-thermal component. For example: NGC4423 and NGC4252 have a reduced chi-squared of $\chi^{2}>2$ and
NGC4241, NGC4301, NGC4206, NGC4466, IC3268, NGC4376 and NGC4222  have $1<\chi^{2}<2$ where $\chi^{2}$ is derived from the best model fit to the data. We nevertheless use the best single temperature fitting parameters for these galaxies and defer more complicated fits to later papers.

Derived dust masses and temperatures (75 galaxies) within these 500 $\mu$m determined apertures are in the range of $10^{6.22-8.17}$ M$_{\odot}$ and 12.8-27.2K respectively with mean values of $10^{7.31}$ M$_{\odot}$ and 20.0K (Table 4). This illustrates the potential sensitivity of the full depth survey to low dust masses ($<10^{6}$ M$_{\odot}$) and the existence of a significant cold dust component (T$<20$K). Prior to the availability of observations at wavelengths longer than about 100 $\mu$m, calculated galaxy dust masses and temperatures from surveys similar to HeViCS were typically $10^{6.6}$ M$_{\odot}$ and 30-50K (taken from Soifer et al., 1987 where they have 31 galaxies in common with HeViCS, see also Devereux and Young, 1990). The typical dust mass is now almost an order of magnitude higher and the temperature almost 20K colder (see also Tuffs et al. 2002).


Interestingly, the coldest galaxy in the sample (12.8K) is UGC7557, the galaxy with the most discrepant IRAS 100 $\mu$m flux (Fig. 6). It is barely detected in the PACS 100 $\mu$m data, but is relatively strongly detected in the other bands. The data provide a reasonably good fit to a modified blackbody curve with a calculated dust mass of $10^{7.42}$ M$_{\odot}$. UGC7557 is classified as a Scd galaxy and appears to be of rather low surface brightness in the optical. Compared to the other galaxies in this sample it is also a little under-luminous in the optical with a B band absolute magnitude of -18.5. If it is very cold dust then its origin is unclear, being apparently of low surface brightness, the dust may just be further away from the stars. Alternately, the dust in this low luminosity galaxy may have a different emissivity and/or size distribution. We will be investigating  this further once the full depth HeViCS data becomes available. 
The complete distribution of dust temperatures is shown in Fig. 7.

\begin{figure}
\centering
\includegraphics[width=0.47\textwidth]{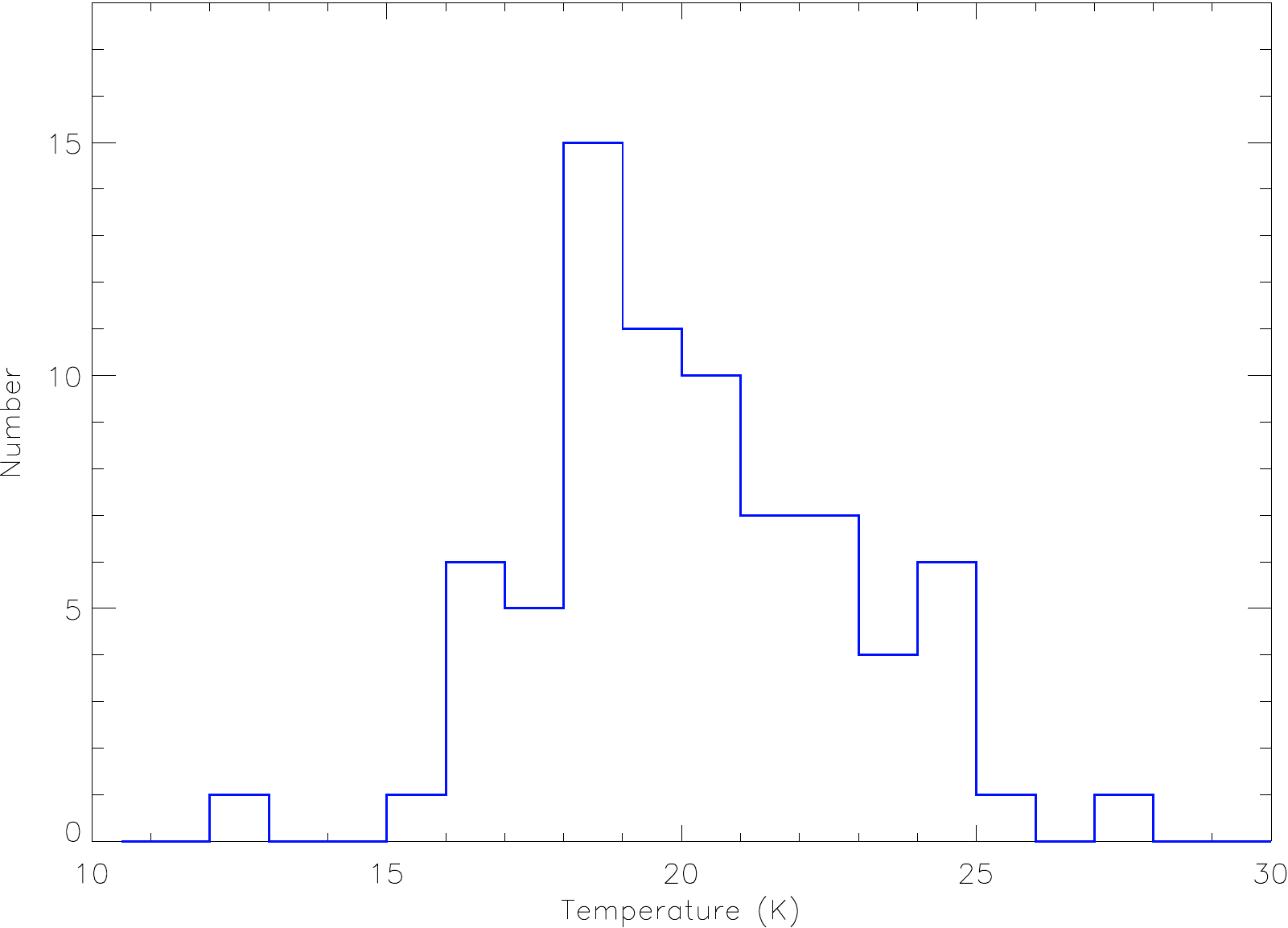}
\caption{The distribution of dust temperature} 
\end{figure}

We have calculated stellar masses ($M_{Star}$) for each galaxy (table 4) using the prescription given in Bell et al. (2003) i.e.
\begin{center}
$\log{M_{star}}=-0.359+0.21(B-V)+\log{\frac{L_{H}}{L_{\odot}}}$
\end{center}
$L_{H}$ has been calculated using a H band absolute magnitude for the Sun of M$^{\odot}_{H}$=3.32 and (B-V) values have been taken from the GOLDMINE database (Gavazzi et al., 2003).

Where available (71 out of 78 galaxies), we have also taken atomic hydrogen gas masses ($M_{HI}$) for each galaxy from the GOLDMINE database, these are also given in table 4 along with the dust masses. In Fig. 8 we show the distribution of mass in each of the three components, dust, gas and stars. In Fig. 9 we show the mass ratio of stars-to-atomic gas and atomic gas-to-dust, these have mean values of 15.1 and 58.2 respectively. We suggest that a factor of $\sim2.9$ (see below) can be used to convert atomic gas to total gas (HI+H$_{2}$) in which case the mean values of the mass ratios are stars-to-gas $\sim5$ and gas-to-dust $\sim170$. The local value of the gas-to-dust ratio for the Milky Way is $\sim143$ (Draine et al., 2007).

\begin{figure}
\centering
\includegraphics[width=0.47\textwidth]{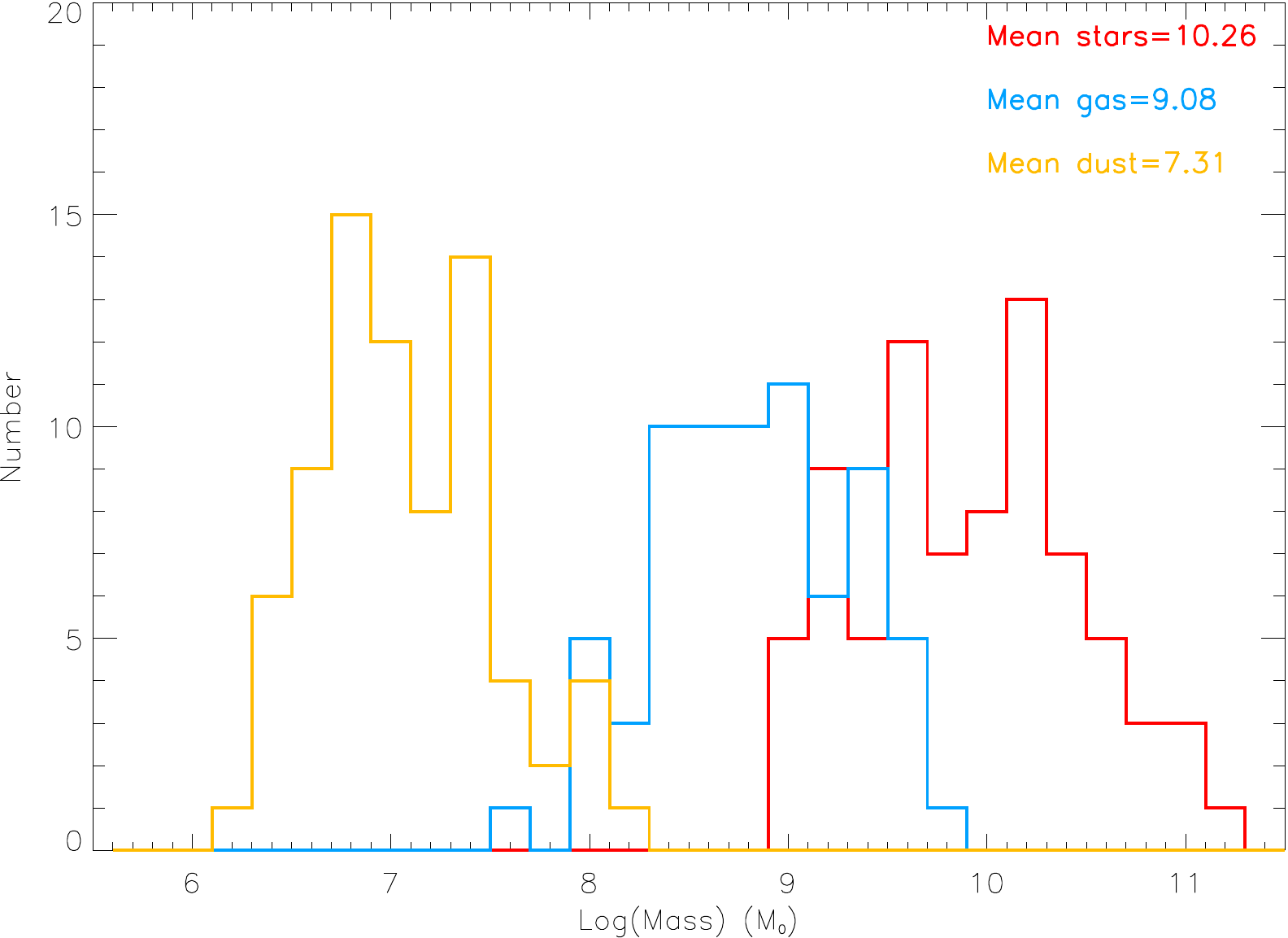}
\caption{The distribution of stellar mass, atomic gas mass and dust mass. All 78 galaxies have measured stellar masses, 71 have measured atomic gas masses and 76 measured dust masses.} 
\end{figure}

\begin{figure}
\centering
\includegraphics[width=0.47\textwidth]{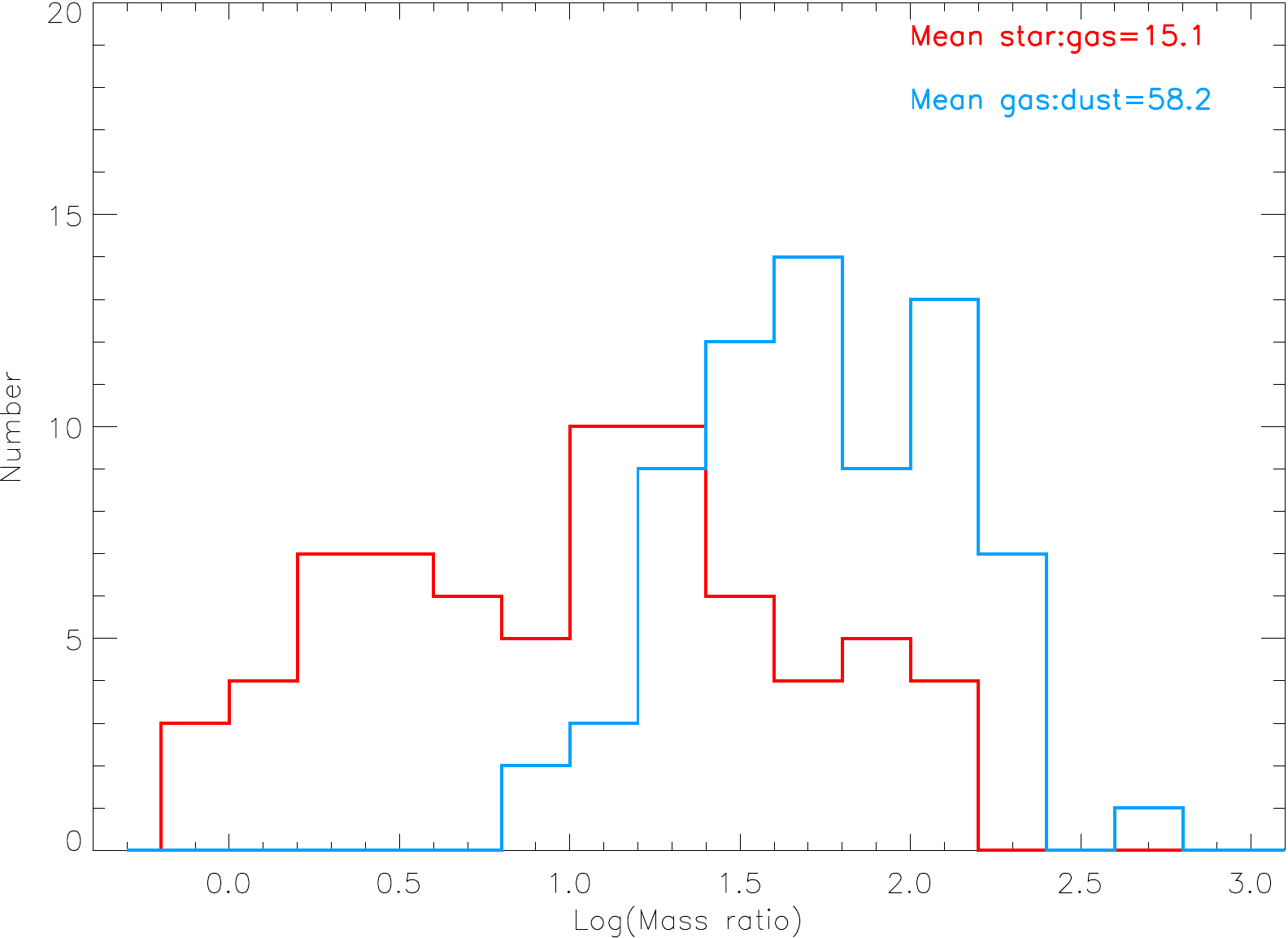}
\caption{The stars-to-atomic gas mass ratio (red) and atomic gas-to-dust mass ratio (blue). Numbers quoted are the mean of the individual values for each galaxy.} 
\end{figure}

By very simply dividing the total mass in each component by the volume sampled ($\sim$181.1(47.2) Mpc$^{3}$) we can calculate the cluster mass densities. Provided the luminosity distributions in each far-infrared band are peaked the total cluster dust mass should be well constrained by our sample galaxies unless there is a significant dust component in the inter-galactic medium (Stickel, et al. 1998). From the dust in our sample galaxies we derive a cluster dust mass density of $8.6(27.8)\pm 1.6 \times 10^{6}$ M$_{\odot}$ Mpc$^{-3}$. This compares with a recent determination of the local dust mass density for galaxies in all environments of $2.2\pm 0.4 \times 10^{5}$ M$_{\odot}$ Mpc$^{-3}$ (Dunne et al. (2011) i.e. the Virgo cluster is over dense in dust by about a factor of 39(126). The Virgo cluster HI mass function is also peaked (Davies et al. 2004, Taylor 2010), so again the HI gas in these galaxies should provide a good estimate of the cluster total. From our sample of 71 galaxies we derive an HI mass density of $4.6(13.9) \times 10^{8}$ M$_{\odot}$ Mpc$^{-3}$. The error on this is uncertain because the HI masses come from a wide variety of sources. Recently, Davies et al. (2011) measured a local 'field' HI mass density of $7.9\pm1.2 \times 10^{7}$ M$_{\odot}$ Mpc$^{-3}$ (see also Martin et al. 2010). Thus the Virgo cluster is over dense in HI by a factor of only about 6(18). Using the Davies et al. and Dunne et al. values for the global HI and dust mass densities gives a local 'field' atomic gas to dust ratio of 359, while the same ratio in the cluster is only 54(50) (total atomic gas mass divided by total dust mass). 
At face value these relative over densities in dust and gas indicate that the cluster galaxies have either processed more gas through stars, and hence created more metals, or they have lost gas and not dust when compared to the general population of galaxies. HI complexes within the cluster, but external to previously identified galaxies, have been found by Kent et al. (2007), but they only represent a small fraction of the cluster atomic gas. 

Comparing stellar mass functions is not quite so straight forward because of uncertainties in the faint end slope of the Virgo cluster luminosity distribution i.e the numbers of faint dwarf galaxies (see Sabatini et al., 2003, and references therein). We obtain from the stars in our sample galaxies a value of $7.8(29.7) \times 10^{9}$ M$_{\odot}$ Mpc$^{-3}$ for the stellar density. Baldry et al. (2008) recently derived the local galaxy stellar mass function and from it a stellar mass density of $2.3\pm1.2 \times 10^{8}$ M$_{\odot}$ Mpc$^{-3}$. This is 34(129) times smaller than our value for Virgo. Interestingly, the stars and dust are in rough agreement as to the value of the over density that Virgo represents, while the atomic gas does not concur. Where required in the above derivations we have used H$_{0}$=72 km s$^{-1}$, $\Omega_{m}=0.27$, $\Omega_{\Lambda}=0.73$.

\section{Conclusions}
The Herschel Virgo Cluster Survey is providing unprecedented high resolution data, at far-infrared wavelengths, on some of the largest angular sized galaxies in the sky. In this paper we have described some of the global properties of these galaxies. The Virgo bright galaxy sample discussed here consists of 78 galaxies selected at 500 $\mu$m each with a confirmed Virgo cluster optical counterpart. Each galaxy is detected in each of the five (100-500 $\mu$m) far-infrared bands. We have carried out aperture photometry on each galaxy and compared, where possible, our data with that previously produced by IRAS, ISO, Spitzer and Planck. Table 1 gives a complete list of the galaxies in the sample and their flux densities in each far-infrared band. 

These flux densities enable us to make a measurement of the luminosity function/distribution in each band. These luminosity function/distributions are not power laws but 'peaked' functions. If subsequent measurements of the luminosity distributions in more wide-ranging environments have the more common power law form (Schechter) then this is a clear indication of environmental processes in the cluster affecting the far-infrared properties of galaxies.

We define and measure for each galaxy a 100-500 $\mu$m far-infrared and an optical 0.4-2.5 $\mu$m luminosity ($L_{100-500}$ and $L_{0.4-2.5}$). Assuming a simple screen model of dust these luminosities can be used to calculate a 'typical' optical depth ($<\tau>$) for photons emerging from a galaxy. Values of $<\tau>$ range from zero to an optically thick 2.8. With a mean value of $<\tau_{mean}>=$0.4 most cluster (stellar) photons are emerging from reasonably optically thin optical depths. We also use the values of $L_{100-500}$ and $L_{0.4-2.5}$ for each galaxy to calculate far-infrared and optical luminosity densities of $1.6(7.0)\pm0.2\times10^{9}$ L$_{\odot}$ Mpc$^{-3}$ and $5.0(20.0)\times10^{9}$ L$_{\odot}$ Mpc$^{-3}$ respectively. The optical luminosity density is some 3.2(2.9) times larger than the far-infrared.

The derived flux densities in each band are also used to calculate a dust mass and temperature for each galaxy assuming a single component modified blackbody with $\beta=2$. The mean (typical) dust mass and temperature for the galaxies in this sample are $\log{(M_{Dust})}=7.31$ and $T_{D}=20.0$K respectively. We have also used data from the literature to calculate both stellar and atomic gas masses for each galaxy, these have mean values of 
$\log{(M_{Stars})}=10.26$ and $\log{(M_{HI})}=9.08$ respectively. The total masses in each of these baryonic components are used to calculate mass densities of $8.6(27.8) \times10^{6}$, $4.6(13.9) \times10^{8}$, $7.8(29.7) \times10^{9}$ M$_{\odot}$ Mpc$^{-3}$ respectively for cluster dust, atomic gas and stars. These values are higher than that derived using galaxies in the general field by factors of 39(126), 6(18) and 34(129) respectively. Atomic gas appears to have been lost or consumed more effectively than is typical in field galaxies.

The above information on the properties of Virgo cluster galaxies and on the properties of the cluster as a whole, will hopefully be useful when compared to the global properties of galaxies throughout the local Universe and also with studies of other galaxy clusters both near and far.

\vspace{0.5cm}
\noindent
{\bf ACKNOWLEDGEMENTS} \\

The Herschel spacecraft was designed, built, tested, and launched under a contract to ESA managed by the Herschel/Planck Project team by an industrial consortium under the overall responsibility of the prime contractor Thales Alenia Space (Cannes), and including Astrium (Friedrichshafen) responsible for the payload module and for system testing at spacecraft level, Thales Alenia Space (Turin) responsible for the service module, and Astrium (Toulouse) responsible for the telescope, with in excess of a hundred subcontractors.

PACS has been developed by a consortium of institutes led by MPE (Germany) and including UVIE (Austria); KU Leuven, CSL, IMEC (Belgium); CEA, LAM (France); MPIA (Germany); INAF-IFSI/OAA/OAP/OAT, LENS, SISSA (Italy); IAC (Spain). This development has been supported by the funding agencies BMVIT (Austria), ESA-PRODEX (Belgium), CEA/CNES (France), DLR (Germany), ASI/INAF (Italy), and CICYT/MCYT (Spain).

SPIRE has been developed by a consortium of institutes led by Cardiff University (UK) and including Univ. Lethbridge (Canada); NAOC (China); CEA, LAM (France); IFSI, Univ. Padua (Italy); IAC (Spain); Stockholm Observatory (Sweden); Imperial College London, RAL, UCL-MSSL, UKATC, Univ. Sussex (UK); and Caltech, JPL, NHSC, Univ. Colorado (USA). This development has been supported by national funding agencies: CSA (Canada); NAOC (China); CEA, CNES, CNRS (France); ASI (Italy); MCINN (Spain); SNSB (Sweden); STFC (UK); and NASA (USA).

This research has made use of the NASA/IPAC Extragalactic Database (NED) which is operated by the Jet Propulsion Laboratory, California Institute of Technology, under contract with the National Aeronautics and Space Administration. 

Funding for the SDSS and SDSS-II has been provided by the Alfred P. Sloan Foundation, the Participating Institutions, the National Science Foundation, the U.S. Department of Energy, the National Aeronautics and Space Administration, the Japanese Monbukagakusho, the Max Planck Society, and the Higher Education Funding Council for England. The SDSS Web Site is http://www.sdss.org/.
The SDSS is managed by the Astrophysical Research Consortium for the Participating Institutions. The Participating Institutions are the American Museum of Natural History, Astrophysical Institute Potsdam, University of Basel, University of Cambridge, Case Western Reserve University, University of Chicago, Drexel University, Fermilab, the Institute for Advanced Study, the Japan Participation Group, Johns Hopkins University, the Joint Institute for Nuclear Astrophysics, the Kavli Institute for Particle Astrophysics and Cosmology, the Korean Scientist Group, the Chinese Academy of Sciences (LAMOST), Los Alamos National Laboratory, the Max-Planck-Institute for Astronomy (MPIA), the Max-Planck-Institute for Astrophysics (MPA), New Mexico State University, Ohio State University, University of Pittsburgh, University of Portsmouth, Princeton University, the United States Naval Observatory, and the University of Washington.

This publication makes use of data products from the Two Micron All Sky Survey, which is a joint project of the University of Massachusetts and the Infrared Processing and Analysis Center/California Institute of Technology, funded by the National Aeronautics and Space Administration and the National Science Foundation.

S. B., L. M., C. P., L. H., S. D. A., E. C., G. G. acknowledge financial support by ASI through the ASI-INAF grant "HeViCS: the Herschel Virgo Cluster Survey" I/009/10/0.

This work received support from the ALMA-CONICYT Fund for the
Development of Chilean Astronomy (Project 31090013) and from the
Center of Excellence in Astrophysics and Associated Technologies (PBF
06).

\vspace{0.5cm}
\noindent
\large
{\bf References} \\
\small
Abazajian K. et al., ApJS, 182, 543  \\
Alton P., Trewhella M., Davies J., Evans R., Bianchi S., Gear W., Thronson H., Valentijn E. and Witt A.,  1998, A\&A, 335, 807 \\
Alton P., Bianchi S., Rand R., Xilouris E., Davies J. and Trewhella M., 1998b, ApJ, 507, L125  \\
Babbedge et al., 2006, MNRAS, 370, 1159 \\
Baes M., Clemens M., Xilouris E., et al., 2010, A\&A, 518, 53 \\
Baes M. and Dejonghe H., 2001, MNRAS, 326, 733 \\
Baldry K., Glazebrook K. and Driver S., 2008, MNRAS, 388, 945  \\
Bell E., McIntosh Daniel H., Katz N. and Weinberg M., 2003, ApJS, 149, 289 \\
Bendo G. et al., 2002, AJ, 123, 3067  \\
Bendo G. et al., 2003, AJ, 125, 2361 \\
Bendo G. et al., 2010, A\&A, 518, 65  \\
Bendo G. et al., 2011, astro-ph/1109.0237 \\ 
Bertin E. and Arnouts S., 1996, A\&ASS, 117, 393 \\
Bianchi S., Davies J., Alton P., Gerin M. and Casoli F., 2000, A\&A, 353, 13 \\
Bianchi S., Davies J. and Alton P., 2000a, A\&A, 359, 65 \\
Bianchi S., 2008, A\&A, 490, 461 \\
Bicay M. and Giovanelli R., 1987, ApJ, 321, 645  \\
Binggeli B., et al., 1985, AJ, 90, 1681  \\
Binggeli B. Tammann G. and Sandage A., 1987, AJ, 94, 251  \\
Binggeli B., Popescu C. and Tammann G., 1993, A\&ASS, 98, 275  \\
Boehringer et al., 1994, Nature, 368, 828  \\
Boselli et al., 2011, A\&A, 528, 107  \\
Boselli et al., 2010, A\&A, 518, 61 \\
Boselli A., Lequeux J. and Gavazzi G., 2002, A\&A, 384, 33 \\
Boselli A. and Gavazzi G., 2006, PASP, 118, 517 \\
Calzetti, D. et al., 2010, ApJ, 714, 1256 \\
Chung, A., van Gorkom J., Kenney J., Crowl H. and Vollmer B., AJ, 139, 1741 \\
Clemens M., et al., 2010, A\&A, 518, 50  \\
Corbelli et al., 2011, A\&A, submitted \\
Cortese L., Davies J., Baes M., et al., 2010, A\&A, 518, 49 \\
Cortese L., Catinella B., Boissier S., Boselli A. and Heinis S., 2011, arXiv:1103.5889 \\
Dame T., Hartmann D., and Thaddeus, P., 2001, ApJ, 547, 792  \\
Davies J., et al., 2010, A\&A, 518, 48 \\
Davies J., et al., 2004, MNRAS, 349, 922 \\
Davies J., Alton P., Trewhella M., Evans R. and Bianchi S.,  1999, MNRAS, 304, 495  \\
Devereux and Young, 1990, ApJ, 359, 42  \\
De Looze I., et al., 2010, A\&A, 518, 54  \\
Dowell C., et al., 2010, Proc. SPIE 7731, 773136 \\
Doyon R. and Joseph R., 1989, MNRAS, 239, 347  \\
Draine B., 2003 ARA\&A, 41, 241  \\
Draine B., et al., 2007, ApJ, 663, 866 \\
Doyon R. and Joseph R., MNRAS, 1989, 239, 347  \\
Dunne L. et al., 2000, MNRAS, 315, 115 \\
Dunne L. and Eales S.,  2002, Ap\&SS, 281, 321 \\
Dunne L. et al., 2010,  arXiv1012.5186 \\
Dwek E., 1998, ApJ, 501, 643 \\
Eales et al., 2009, ApJ, 707, 1779 \\
Eales S., et al., 2010, PASP, 122, 499 \\
Edmunds M., 1990, MNRAS, 246, 678 \\
Edmunds M. and Eales S., 1998, MNRAS, 299, L29 \\
Galametz M. et al., 2009, A\&A, 508, 645 \\
Gavazzi G., Boselli A., Scodeggio M., Pierini D. and Belsole E., 1999, MNRAS, 304, 595 \\
Gavazzi, G. Boselli, A. Donati, A. Franzetti, P. and Scodeggio, M., 2003, A\&A, 400, 451  \\
Giovanelli, R. and Haynes M., 1985, ApJ, 292, 404 \\
Giovanelli, R., et al. 2005, AJ, 130, 2598 \\
Girardi M., Giuricin G., Mardirossian F., Mezzetti M. and Boschin W., 1998, ApJ, 505 74  \\
Gomez et al., 2003, ApJ., 584, 210 \\
Griffin et al., 2010, A\&A, 518, 3 \\
Griffin et al., 2009, EAS, 34, 33G \\
Grossi et al., 2010, A\&A, 518, 52  \\
Haynes M. and Giovanelli R., 1984, AJ, 89, 758 \\
Ibar et al., 2010, MNRAS, 409, 38  \\
Kennicutt, R. C., Jr., et al., 2003, PASP, 115, 928  \\
Kent et al., 2007, ApJ, 665, 15  \\
Lewis I., et al., 2002, MNRAS, 334, 673 \\
Magrini, L., et al., 2011, A\&A, accepted (arXiv:1106.0618) \\
Martin A., Papastergis E., Giovanelli R., Haynes M., Springob, C. and Stierwalt S., 2010, ApJ, submitted (arXiv:1008.5107) \\
Mei S., et al., 2007, ApJ, 655, 144  \\
Meyer D., Jura M. and Cardelli J., 1998, ApJ, 493, 222 \\
Mei  S., et al., 2010, Bulletin of the American Astronomical Society, Vol. 42, p.514
Neugebauer G. et al., 1984, ApJ, 278, L1  \\
Ott S., 2010, ASPC, 434, 139O \\
Pilbratt et al., 2010, A\&A, 518, 1 \\
Poglitsch et al., 2010, A\&A, 518, 2 \\
Pohlen et al., 2010, A\&A, 518, 72 \\
Popescu C., Tuffs R., Volk H., Pierini D. and Madore B., 2002, AJ, 567, 221  \\
Popescu C. and Tuffs R., 2002, MNRAS, 335, L41  \\
Rines K. and Diaferio A., 2006, AJ, 132, 1275  \\
Rodighiero et al., 2010, A\&A, 515, 8 \\
Rowan-Robinson M., Helou G. and Walker D., 1987, MNRAS, 227, 589   \\
Sabatini et al.,  2003, MNRAS, 341, 981  \\
Sandage A., Binggeli B. and Tammann G., 1985, AJ, 90, 1759  \\
Saunders et al., 1990, MNRAS, 242, 318  \\
Serjeant et al., 2001, MNRAS, 322, 262 \\
Skillman E., Kennicutt R., Shields G. and Zaritsky D., 1996, ApJ, 462, 147 \\
Skrutskie M., et al., 2006, AJ, 131, 1163 \\
Smith M., Vlahakis C., Baes M., et al., 2010, A\&A, 518, 51 \\
Soifer et al., 1987, ApJ, 320, 238 \\
Stickel M., Lemke D., Mattila K., Haikala L. and Haas M., 1998, A\&A, 329, 55  \\
Stickel M., Lemke D., Klaas U., Krause O. and Egner S., 2004, A\&A, 422, 39S  \\
Stierwalt S., 2010, ApJ, 723, 1359 \\
Swinyard et al., 2010, A\&A, 518, 4 \\
Takeuchi T., Ishii T., Dole H., Dennefeld M., Lagache, G. and Puget J., 2006, A\&A, 448, 525 \\
Taylor R., 2010, PhD thesis, Cardiff University, UK. \\
Trewhella M., Davies J., Alton P., Bianchi S. and Madore B., 2000, ApJ, 543, 153  \\
Tuffs R. et al., 2002, ApJS, 139, 37  \\
Valhakis C., Dunne L. and Eales S., 2005, MNRAS, 364, 1253 \\
Vila-Costas M. and Edmunds M., 1992, MNRAS, 259, 121  \\
Warren S. et al., 2007, MNRAS, 375, 213 \\
Whittet D., 1991, 'Dust in the galactic Environment', IOP publishing, Bristol. \\

\setcounter{table}{0}
\onecolumn
\begin{center}
\begin{longtable}{cccccccccc} \hline
(1) & (2) & (3) & (4) & (5) & & & (6) & & \\ 
Name    &  RA        &   Dec    & $v$           & D     & $F_{500}  $ & $F_{350}$  & $F_{250}$  & $F_{160}$  & $F_{100}$ \\ 
        & (J2000)    &  (J2000) & (km s$^{-1}$) & (Mpc) & (Jy) & (Jy) & (Jy) & (Jy) & (Jy) \\ \hline
NGC4165 & 12:12:12.6 &  13:14:40.0 &  1862 & 32.0 &  0.15 &  0.38 &  0.81 &  1.19 &  0.61 \\
IC00769 & 12:12:32.6 &  12:07:22.1 &  2209 & 32.0 &  0.40 &  0.96 &  1.81 &  3.04 &  0.99 \\
NGC4189 & 12:13:47.8 &  13:25:32.1 &  2114 & 32.0 &  0.99 &  2.88 &  6.91 & 12.26 & 11.34 \\
NGC4192 & 12:13:48.4 &  14:54:00.5 &  -139 & 17.0 &  5.07 & 12.87 & 27.75 & 37.65 & 24.24 \\
NGC4193 & 12:13:53.7 &  13:10:22.0 &  2470 & 32.0 &  0.68 &  1.68 &  3.51 &  6.93 &  2.62 \\
NGC4197 & 12:14:38.8 &  05:48:23.0 &  2062 & 32.0 &  0.67 &  1.68 &  3.37 &  6.37 &  6.32 \\
IC03061 & 12:15:04.6 &  14:01:42.9 &  2316 & 17.0 &  0.32 &  0.77 &  1.56 &  2.35 &  1.68 \\
NGC4206 & 12:15:17.1 &  13:01:28.4 & 704   & 17.0 &    1.08 &  2.20 &  3.56 &  4.69 &  2.35 \\
NGC4212 & 12:15:39.0 &  13:54:08.7 & -88   & 17.0 &    1.85 &  5.33 & 12.96 & 26.77 & 21.78 \\
NGC4216 & 12:15:54.9 &  13:08:52.7 & 139   & 17.0 &    3.93 &  9.99 & 21.10 & 31.89 & 16.47 \\
NGC4222 & 12:16:22.5 &  13:18:20.9 & 229   & 17.0 &    0.86 &  1.95 &  3.35 &  4.19 &  4.02 \\
NGC4234 & 12:17:09.2 &  03:40:51.1 &  2031 & 32.0 &  0.31 &  0.93 &  2.11 &  4.80 &  4.73 \\
NGC4237 & 12:17:11.5 &  15:19:26.6 & 865   & 17.0 &    1.01 &  3.00 &  7.38 & 15.16 & 10.81  \\
NGC4241 & 12:17:25.6 &  06:41:19.6 &  2239 & 32.0 &  0.23 &  0.59 &  1.17 &  1.86 &  0.44 \\
IC03115 & 12:18:00.1 &  06:39:00.2 & 732   & 23.0 &    0.32 &  0.62 &  1.03 &  1.25 &  0.38 \\
NGC4252 & 12:18:31.6 &  05:33:53.6 & 864   & 32.0 &    0.12 &  0.24 &  0.35 &  0.55 &  0.61 \\
NGC4254 & 12:18:49.7 &  14:25:10.8 &  2404 & 17.0 &  8.79 & 26.29 & 65.45 & 130.99 & 114.25 \\
NGC4260 & 12:19:21.2 &  06:06:03.2 &  1935 & 23.0 &  0.16 &  0.44 &  0.88 &  1.33 &  0.68 \\
NGC4266 & 12:19:42.7 &  05:32:18.6 &  1617 & 32.0 &  0.17 &  0.53 &  1.30 &  2.63 &  1.64 \\
NGC4273 & 12:19:56.1 &  05:20:36.3 &  2379 & 32.0 &  1.47 &  4.20 & 10.45 & 24.31 & 26.01 \\
UGC7387 & 12:20:17.8 &  04:12:04.8 &  1732 & 17.0 &  0.19 &  0.48 &  0.94 &  1.52 &  0.87 \\
NGC4289 & 12:21:02.9 &  03:43:24.7 &  2541 & 17.0 &  0.48 &  1.22 &  2.29 &  3.14 &  2.31 \\
NGC4294 & 12:21:17.8 &  11:30:37.8 & 355   & 17.0 &    0.89 &  2.06 &  4.16 &  7.27 &  5.78 \\
NGC4298 & 12:21:33.3 &  14:36:15.2 &  1136 & 17.0 &  1.66 &  4.94 & 11.89 & 21.95 & 15.62 \\
NGC4299 & 12:21:40.7 &  11:30:02.6 & 232   & 17.0 &    0.46 &  1.16 &  2.41 &  4.75 &  4.83  \\
NGC4302 & 12:21:42.4 &  14:35:50.4 &  1150 & 17.0 &  2.89 &  8.24 & 18.67 & 30.44 & 18.73 \\
NGC4303 & 12:21:55.1 &  04:28:27.8 &  1566 & 17.0 &  8.08 & 22.55 & 54.57 & 114.90 & 102.95 \\
NGC4307 & 12:22:05.7 &  09:02:29.0 &  1035 & 23.0 &  0.72 &  2.06 &  4.58 &  7.75 &  4.71 \\
NGC4309 & 12:22:12.1 &  07:08:43.9 &  1071 & 23.0 &  0.22 &  0.55 &  1.10 &  2.03 &  1.70 \\
NGC4301 & 12:22:27.6 &  04:33:49.3 &  1278 & 17.0 &  0.17 &  0.45 &  0.81 &  1.44 &  1.52 \\
NGC4312 & 12:22:31.3 &  15:32:19.0 & 148   & 17.0 &    0.55 &  1.66 &  4.19 &  8.62 &  7.33 \\
NGC4313 & 12:22:38.4 &  11:48:01.5 &  1442 & 17.0 &  0.62 &  1.81 &  4.11 &  8.23 &  4.73 \\
IC03225 & 12:22:38.7 &  06:40:37.3 &  2362 & 23.0 &  0.20 &  0.46 &  0.90 &  1.40 &  0.81 \\
NGC4316 & 12:22:42.0 &  09:20:00.0 &  1250 & 23.0 &  0.84 &  2.16 &  4.91 &  9.12 &  5.92 \\
NGC4321 & 12:22:54.9 &  15:49:24.8 &  1575 & 17.0 &  9.98 & 28.54 & 68.31 & 109.45 & 76.34 \\
NGC4324 & 12:23:06.6 &  05:15:03.8 &  1668 & 17.0 &  0.36 &  0.95 &  1.94 &  2.71 &  1.69 \\
NGC4330 & 12:23:17.2 &  11:22:02.0 &  1563 & 17.0 &  0.74 &  1.80 &  3.54 &  5.19 &  2.42  \\
NGC4343 & 12:23:38.3 &  06:57:20.6 &  1013 & 23.0 &  0.58 &  1.59 &  3.60 &  6.70 &  5.53 \\
IC03258 & 12:23:43.7 &  12:28:40.7 &  -430 & 17.0 &  0.17 &  0.40 &  0.72 &  1.37 &  0.80 \\
IC03259 & 12:23:47.3 &  07:11:23.8 &  1420 & 23.0 &  0.22 &  0.53 &  1.06 &  1.37 &  0.95 \\
NGC4351 & 12:24:01.5 &  12:12:17.4 &  2316 & 17.0 &  0.26 &  0.68 &  1.41 &  2.65 &  1.41 \\
IC03268 & 12:24:07.7 &  06:36:22.9 & 727   & 23.0 &    0.17 &  0.41 &  0.79 &  1.50 &  1.77 \\
NGC4374 & 12:25:03.7 &  12:52:55.4 & 910   & 17.0 &    0.13 &  0.15 &  0.27 &  0.86 &  1.14 \\
NGC4376 & 12:25:18.4 &  05:44:30.1 &  1138 & 23.0 &  0.23 &  0.54 &  1.06 &  2.76 &  1.60 \\
NGC4378 & 12:25:18.5 &  04:55:23.4 &  2557 & 17.0 &  0.56 &  1.85 &  3.36 &  1.10 &  1.00 \\
NGC4380 & 12:25:22.0 &  10:01:06.3 & 963   & 23.0 &    0.76 &  2.23 &  4.91 &  6.17 &  3.04 \\
UGC7513 & 12:25:42.8 &  07:13:01.3 & 993   & 23.0 &    1.02 &  2.45 &  4.91 &  8.35 &  5.87 \\
NGC4388 & 12:25:47.5 &  12:39:44.7 &  2515 & 17.0 &  1.25 &  3.50 &  8.64 & 20.68 & 21.41 \\
NGC4390 & 12:25:51.5 &  10:27:33.3 &  1101 & 23.0 &  0.40 &  0.94 &  1.89 &  2.85 &  1.61  \\
IC03322 & 12:25:54.1 &  07:33:19.5 &  1202 & 23.0 &  0.36 &  0.91 &  1.85 &  2.99 &  2.67 \\
NGC4402 & 12:26:07.4 &  13:06:44.7 & 230   & 17.0 &    2.08 &  6.08 & 14.50 & 28.15 & 21.15 \\
UGC7537 & 12:26:29.6 &  08:52:19.8 &  1278 & 23.0 &  0.26 &  0.60 &  1.11 &  1.58 &  0.94 \\
NGC4413 & 12:26:32.3 &  12:36:38.2 & 103   & 17.0 &    0.44 &  1.16 &  2.50 &  4.36 &  1.94 \\
NGC4412 & 12:26:36.5 &  03:57:55.6 &  2289 & 17.0 &  0.36 &  1.11 &  2.77 &  6.60 &  6.88 \\
NGC4416 & 12:26:46.8 &  07:55:06.8 &  1390 & 17.0 &  0.40 &  1.08 &  2.39 &  4.79 &  2.93 \\
UGC7546 & 12:26:47.6 &  08:52:54.2 &  1272 & 23.0 &  0.56 &  1.24 &  2.25 &  2.62 &  1.08 \\
NGC4423 & 12:27:09.8 &  05:52:56.5 &  1120 & 23.0 &  0.30 &  0.59 &  0.99 &  1.93 &  1.15 \\
UGC7557 & 12:27:10.4 &  07:15:58.7 & 932   & 23.0 &    0.36 &  0.73 &  1.14 &  0.74 &  0.13 \\
NGC4424 & 12:27:11.3 &  09:25:14.5 & 437   & 23.0 &    0.32 &  0.96 &  2.47 &  5.52 &  6.21 \\
NGC4435 & 12:27:40.3 &  13:04:43.9 & 775   & 17.0 &    0.21 &  0.68 &  1.87 &  4.31 &  4.77 \\
NGC4438 & 12:27:44.7 &  13:00:25.5 & 104   & 17.0 &    1.07 &  3.23 &  7.62 & 15.54 & 11.40  \\
NGC4445 & 12:28:15.9 &  09:26:10.4 & 328   & 23.0 &    0.20 &  0.59 &  1.39 &  2.44 &  1.73 \\
NGC4451 & 12:28:40.6 &  09:15:40.7 & 858   & 23.0 &    0.30 &  0.87 &  2.21 &  5.24 &  4.90 \\
NGC4459 & 12:29:00.1 &  13:58:46.0 &  1210 & 17.0 &  0.19 &  0.61 &  1.63 &  4.26 &  5.20 \\
NGC4469 & 12:29:28.1 &  08:44:58.7 & 508   & 23.0 &    0.25 &  0.84 &  2.05 &  4.02 &  3.80 \\
NGC4466 & 12:29:29.9 &  07:41:43.1 & 759   & 17.0 &    0.19 &  0.40 &  0.83 &  1.13 &  1.10 \\
NGC4470 & 12:29:37.9 &  07:49:26.9 &  2341 & 17.0 &  0.36 &  1.04 &  2.44 &  5.84 &  5.26 \\
NGC4486 & 12:30:49.2 &  12:23:24.8 &  1292 & 17.0 &  1.28 &  0.92 &  0.67 &  0.92 &  0.53 \\
NGC4492 & 12:30:59.6 &  08:04:35.2 &  1777 & 17.0 &  0.22 &  0.65 &  1.37 &  2.39 &  1.07 \\
IC03476 & 12:32:40.9 &  14:02:52.3 &  -170 & 17.0 &  0.43 &  1.09 &  2.19 &  3.34 &  3.41 \\
NGC4519 & 12:33:30.5 &  08:39:17.6 &  1216 & 17.0 &  1.03 &  2.47 &  5.06 &  9.44 &  6.48 \\
NGC4522 & 12:33:39.7 &  09:10:32.4 &  2329 & 17.0 &  0.65 &  1.68 &  3.56 &  6.29 &  5.12 \\
NGC4526 & 12:34:03.1 &  07:41:57.3 & 448   & 17.0 &    0.96 &  3.02 &  7.90 & 17.59 & 17.33  \\
NGC4531 & 12:34:16.0 &  13:04:32.2 & 195   & 17.0 &    0.22 &  0.63 &  1.44 &  2.59 &  1.15 \\
NGC4535 & 12:34:20.1 &  08:11:53.5 &  1964 & 17.0 &  5.81 & 15.63 & 34.43 & 49.42 & 21.89 \\
IC03521 & 12:34:39.3 &  07:09:43.2 & 593   & 17.0 &    0.18 &  0.60 &  1.48 &  2.93 &  2.36 \\
NGC4567 & 12:36:32.4 &  11:15:27.0 &  2277 & 17.0 &  1.68 &  5.01 & 12.64 & 26.49 & 20.76 \\
NGC4568 & 12:36:33.9 &  11:14:35.1 &  2255 & 17.0 &  4.12 & 12.33 & 30.89 & 64.45 & 55.58 \\ \hline
\caption{The Herschel Virgo Cluster Survey Bright Galaxy Sample - (1) name, (2) (3) position, (4) velocity, (5) distance, and (6) far-infrared flux density. Note that NGC4567 and NGC4568 are very close together and so it is difficult to separate the measurements of one from the other.}
\end{longtable}
\end{center}

\setcounter{table}{3}
\begin{center}
\begin{longtable}{ccccccccccc} \hline
(1) & (2) & (3) & (4) & (5) & (6) & (7) & (8) & (9) & (10) & (11)\\ 
Name    & M$_{B}$ & Log(M$_{Stars}$)  & Log(M$_{HI}$)  & Log(M$_{Dust}$)  & T$_{d}$ 
&  Log(L$_{0.4-2.5}$) & Log(L$_{100-500}$) & $<\tau>$ & M$_{Stars}$/M$_{HI}$ & M$_{HI}$/M$_{Dust}$ \\
       &          &    (M$_{\odot}$)  & (M$_{\odot}$)  & (M$_{\odot}$)    & (K)
&    (L$_{\odot}$) & (L$_{\odot}$)  & & &  \\ \hline           
NGC4165 & -17.99 &  9.77 & 8.39 &  6.97$\pm$0.07    & 18.2$\pm$0.9 & 9.67 & 8.93 & 0.17  & 24.0&  26.3\\
IC00769 & -19.13 &  9.98 & 9.48 &  7.44$\pm$0.07    & 17.0$\pm$0.8 & 9.71 & 9.26 & 0.30 & 3.2& 109.6\\
NGC4189 & -19.84 & 10.39 &  9.40 & 7.65$\pm$0.07   &  21.8$\pm$1.3 &10.08 & 10.05 & 0.66 & 9.8&  56.2\\
NGC4192 & -20.42 & 10.88 &  9.63 & 7.93$\pm$0.07   &  18.4$\pm$0.9 &10.56 & 9.93 & 0.21 & 17.8&  50.1\\
NGC4193 & -19.24 & 10.29 &  9.24 & 7.60$\pm$0.07   &  18.4$\pm$0.9 &10.11 & 9.62 & 0.28 & 11.2&  43.7\\
NGC4197 & -19.15 &  9.91 & 9.71 &  7.46$\pm$0.08    & 20.6$\pm$1.3 & 9.80 & 9.79 & 0.68 & 1.6& 177.8\\
IC03061 & -16.77 &  9.23 & 8.79 &  6.69$\pm$0.08    & 18.7$\pm$1.0 & 9.08 & 8.74 & 0.37 & 2.8& 125.9\\
NGC4206 & -18.15 &  9.72 &  9.38 & 7.31$\pm$0.08   &  16.1$\pm$0.8 & 9.63 & 8.99 & 0.21 & 2.2& 117.5\\
NGC4212 & -19.33 & 10.27 &  8.91 & 7.35$\pm$0.07   &  22.4$\pm$1.3 &10.04 & 9.80 & 0.45 & 22.9&  36.3\\
NGC4216 & -20.31 & 10.94 &  9.25 & 7.83$\pm$0.07   &  18.3$\pm$0.9 &10.66 & 9.81 & 0.13 & 49.0&  26.3\\
NGC4222 & -17.20 &  9.57 & 9.03 &  7.20$\pm$0.09    & 16.8$\pm$1.0 & 9.25 & 9.07 & 0.51 & 3.5&  67.6\\
NGC4234 & -19.05 &  9.61 & 8.88 &  7.08$\pm$0.08    & 23.3$\pm$1.5 & 9.36 & 9.65 & 1.09 & 5.4&  63.1\\
NGC4237 & -18.49 & 10.01 &  8.32 & 7.11$\pm$0.07   &  22.0$\pm$1.2 & 9.89 & 9.53 & 0.36 & 49.0&  16.2\\
NGC4241 & -19.55 & 10.28 &  8.45 & 7.28$\pm$0.06   &  16.3$\pm$0.6 &10.21 & 9.01 & 0.06 & 67.6&  14.8\\
IC03115 & -17.96 &  9.35 & 9.03 &  7.11$\pm$0.07    & 15.1$\pm$0.6 & 9.14 & 8.62 & 0.26 & 2.1&  83.2\\
NGC4252 & -17.74 &  9.25 & 9.00 &  6.88$\pm$0.11    & 16.0$\pm$1.1 & 9.26 & 8.77 & 0.28 & 1.8& 131.8\\
NGC4254 & -20.70 & 10.67 &  9.65 & 8.02$\pm$0.07   &  22.7$\pm$1.3 &10.59 & 10.51 & 0.60 & 10.5&  42.7\\
NGC4260 & -19.22 & 10.38 &  - &    6.73$\pm$0.07    & 18.2$\pm$0.9 &10.09 & 8.69 & 0.04 &  - &  -\\
NGC4266 & -19.04 & 10.24 &  - &    6.93$\pm$0.07    & 21.5$\pm$1.1 & 9.81 & 9.29 & 0.26 &  - &  -\\
NGC4273 & -20.04 & 10.33 &  9.54 & 7.71$\pm$0.08   &  24.4$\pm$1.6 &10.19 & 10.38 & 0.93 & 6.2&  67.6\\
UGC7387 & -16.11 &  9.05 & 8.39 &  6.49$\pm$0.07    & 18.5$\pm$0.9 & 8.96 & 8.50 & 0.30 & 4.6&  79.4\\
NGC4289 & -16.56 &  9.52 & 9.02 &  6.92$\pm$0.08    & 17.9$\pm$0.9 & 9.25 & 8.88 & 0.35 & 3.2& 125.9\\
NGC4294 & -18.47 &  9.59 &  9.22 & 7.06$\pm$0.08   &  19.5$\pm$1.1 & 9.59 & 9.24 & 0.37 & 2.3& 144.5\\
NGC4298 & -19.20 & 10.12 &  8.94 & 7.36$\pm$0.07   &  21.3$\pm$1.1 &10.00 & 9.69 & 0.40 & 15.1&  38.0\\
NGC4299 & -18.16 &  8.98 & 9.04 &  6.72$\pm$0.08    & 21.2$\pm$1.4 & 7.94 & 9.11 & 2.76 & 0.9& 208.9\\
NGC4302 & -18.84 & 10.32 &  9.24 & 7.66$\pm$0.07   &  19.7$\pm$1.0 &10.08 & 9.81 & 0.43 & 12.0&  38.0\\
NGC4303 & -20.85 & 10.71 &  9.68 & 7.96$\pm$0.08   &  22.7$\pm$1.4 &10.66 & 10.46 & 0.49 & 10.7&  52.5\\
NGC4307 & -19.03 & 10.12 &  8.15 & 7.31$\pm$0.07   &  19.8$\pm$1.0 &10.09 & 9.48 & 0.22 & 93.3 & 6.9\\
NGC4309 & -18.01 &  9.80 & 7.64 &  6.72$\pm$0.08    & 20.1$\pm$1.2 & 9.65 & 8.96 & 0.19 &144.5 & 8.3\\
NGC4301 & -17.51 &  8.99 & 9.08 &  6.37$\pm$0.09    & 19.6$\pm$1.3 & 9.09 & 8.61 & 0.28 & 0.8& 512.9\\
NGC4312 & -18.69 &  9.94 &  8.08 & 6.82$\pm$0.07   &  22.9$\pm$1.4 & 9.85 & 9.32 & 0.26 & 72.4&  18.2\\
NGC4313 & -18.65 & 10.12 &  8.02 & 6.96$\pm$0.07   &  20.6$\pm$1.1 & 9.93 & 9.22 & 0.18 & 125.9&  11.5\\
IC03225 & -17.32 &  9.16 & 8.81 &  6.76$\pm$0.08    & 18.1$\pm$0.9 & 9.02 & 9.06 & 0.74 & 2.2& 134.9\\
NGC4316 & -18.12 & 10.03 &  9.01 & 7.32$\pm$0.00   &  20.3$\pm$0.0 & 9.81 & 9.55 & 0.44 & 10.5&  49.0\\
NGC4321 & -21.13 & 10.90 &  9.44 & 8.17$\pm$0.07   &  20.3$\pm$1.0 &10.77 & 10.39 & 0.35 & 28.8&  18.6\\
NGC4324 & -18.58 & 10.15 &  8.73 & 6.80$\pm$0.07   &  18.3$\pm$0.9 & 9.94 & 8.78 & 0.07 & 26.3&  85.1\\
NGC4330 & -18.01 &  9.71 &  8.61 & 7.13$\pm$0.07   &  17.5$\pm$0.8 & 9.55 & 9.01 & 0.25 & 12.6&  30.2\\
NGC4343 & -18.61 & 10.16 &  8.78 & 7.13$\pm$0.08   &  21.2$\pm$1.2 & 9.96 & 9.47 & 0.28 & 24.0&  44.7\\
IC03258 & -17.31 &  8.95 & 8.45 &  6.39$\pm$0.08    & 18.6$\pm$1.0 & 7.49 & 8.45 & 2.32 & 3.2& 114.8\\
IC03259 & -17.45 &  9.40 & 8.40 &  6.84$\pm$0.08    & 17.9$\pm$0.9 & 9.23 & 8.77 & 0.30 & 10.0&  33.1\\
NGC4351 & -18.16 &  9.29 & 8.52 &  6.60$\pm$0.07    & 19.3$\pm$1.0 & 9.40 & 8.72 & 0.19 & 5.9&  83.2\\
IC03268 & -18.01 &  9.27 & 8.89 &  6.56$\pm$0.09    & 20.4$\pm$1.4 & 9.35 & 8.92 & 0.31 & 2.4& 213.8\\
NGC4374 & -20.97 & 10.99 & 8.96 &  -                &   -  &10.90 & 8.43 & - & 107.2 & - \\
NGC4376 & -17.98 &  9.35 & 8.83 &  6.69$\pm$0.08    & 20.5$\pm$1.2 & 9.32 & 9.00 & 0.39 & 3.3& 138.0\\
NGC4378 & -18.63 & 10.16 &  8.84 & -                &   -  & 9.96 & 8.65 & 0.05 & 20.9&  - \\
NGC4380 & -19.38 & 10.40 &  8.47 & 7.45$\pm$0.07   &  18.0$\pm$0.8 &10.25 & 9.36 & 0.12 & 85.1&  10.5\\
UGC7513 & -18.04 &  9.91 &  9.45 & 7.42$\pm$0.08   &  19.2$\pm$1.1 & 9.66 & 9.54 & 0.57 & 2.9& 107.2\\
NGC4388 & -19.28 & 10.24 &  8.65 & 7.08$\pm$0.08   &  24.4$\pm$1.6 &10.14 & 9.75 & 0.34 & 38.9&  37.2\\
NGC4390 & -18.39 &  9.51 & 8.90 &  7.08$\pm$0.07    & 18.1$\pm$0.9 & 9.58 & 9.04 & 0.25 & 4.1&  66.1\\
IC03322 & -17.58 &  9.57 & 8.71 &  6.96$\pm$0.08    & 19.6$\pm$1.1 & 9.44 & 9.15 & 0.42 & 7.2&  56.2\\
NGC4402 & -18.51 & 10.11 &  8.66 & 7.43$\pm$0.07   &  21.6$\pm$1.2 & 9.90 & 9.81 & 0.60 & 28.2&  17.0\\
UGC7537 & -18.17 &  9.36 & 9.23 &  6.91$\pm$0.08    & 17.5$\pm$0.9 & 9.15 & 8.80 & 0.37 & 1.3& 208.9\\
NGC4413 & -18.21 &  9.60 &  8.29 & 6.87$\pm$0.07   &  18.7$\pm$0.9 & 9.59 & 8.90 & 0.19 & 20.4&  26.3\\
NGC4412 & -18.06 &  9.59 & 8.33 &  6.56$\pm$0.08    & 24.8$\pm$1.6 & 9.51 & 9.25 & 0.44 & 18.2&  58.9\\
NGC4416 & -17.89 &  9.50 & 8.41 &  6.74$\pm$0.07    & 20.5$\pm$1.1 & 9.36 & 9.00 & 0.36 & 12.3&  46.8\\
UGC7546 & -18.75 &  9.57 & 9.34 &  7.34$\pm$0.07    & 16.0$\pm$0.7 & 9.60 & 8.98 & 0.22 & 1.7& 100.0\\
NGC4423 & -17.92 &  9.23 & 9.20 &  6.85$\pm$0.09    & 18.0$\pm$1.1 & 9.39 & 8.87 & 0.26 & 1.1& 223.9\\
UGC7557 & -18.54 &  9.24 & 9.43 &  7.42$\pm$0.07    & 12.8$\pm$0.4 & 8.83 & 8.45 & 0.35 & 0.6& 102.3\\
NGC4424 & -19.48 & 10.18 &  8.59 & 6.77$\pm$0.07   &  24.8$\pm$1.5 &10.12 & 9.46 & 0.20 & 38.9&  66.1\\
NGC4435 & -19.35 & 10.44 &  - &    6.32$\pm$0.08   &  25.9$\pm$1.7 &10.33 & 9.08 & 0.06 &  - &  -\\
NGC4438 & -20.03 & 10.70 &  8.68 & 7.14$\pm$0.02   &  21.9$\pm$0.3 &10.27 & 9.54 & 0.17 & 104.7&  34.7\\
NGC4445 & -18.12 &  9.86 & 8.06 &  6.72$\pm$0.07    & 20.8$\pm$1.1 & 9.69 & 9.00 & 0.19 & 63.1&  21.9\\
NGC4451 & -18.37 &  9.79 & 8.50 &  6.75$\pm$0.08    & 24.2$\pm$1.5 & 9.34 & 9.39 & 0.75 & 19.5&  56.2\\
NGC4459 & -19.63 & 10.60 &  - &    6.22$\pm$0.08    & 27.2$\pm$2.0 &10.45 & 9.10 & 0.04 &  -&   -\\
NGC4469 & -19.62 & 10.57 &  - &    6.75$\pm$0.07    & 23.2$\pm$1.4 &10.38 & 9.28 & 0.08 &  -&   -\\
NGC4466 & -16.55 &  8.98 & 8.16 &  6.43$\pm$0.08    & 18.4$\pm$1.1 & 8.86 & 8.50 & 0.36 & 6.6&  53.7\\
NGC4470 & -17.97 &  9.43 & 8.63 &  6.58$\pm$0.08    & 23.5$\pm$1.5 & 9.44 & 9.16 & 0.42 & 6.3& 112.2\\
NGC4486 & -21.32 & 11.13 &  -   &  -                &  -   &11.05 & 8.37 &  - &  - &  -\\
NGC4492 & -18.02 &  9.79 &  7.94 & 6.60$\pm$0.07   &  18.8$\pm$0.9 & 9.81 & 8.64 & 0.07 & 70.8&  21.9\\
IC03476 & -17.80 &  9.18 & 8.39 &  6.79$\pm$0.08    & 19.4$\pm$1.2 & 9.07 & 8.97 & 0.59 & 6.2&  39.8\\
NGC4519 & -18.69 &  9.63 &  9.43 & 7.14$\pm$0.08   &  19.7$\pm$1.1 & 9.65 & 9.32 & 0.39 & 1.6& 195.0\\
NGC4522 & -18.18 &  9.57 &  8.63 & 6.94$\pm$0.08   &  20.2$\pm$1.2 & 9.53 & 9.18 & 0.37 & 8.7&  49.0\\
NGC4526 & -20.35 & 10.95 &  9.33 & 7.01$\pm$0.07   &  24.5$\pm$1.5 &10.72 & 9.66 & 0.08 & 41.7& 208.9\\
NGC4531 & -18.68 &  9.99 &  -    & 6.57$\pm$0.07   &  19.2$\pm$0.9 & 9.77 & 8.67 & 0.08 &  - & -\\
NGC4535 & -20.42 & 10.63 &  9.59 & 8.04$\pm$0.07   &  18.0$\pm$0.8 &10.48 & 9.97 & 0.27 & 11.0&  35.5\\
IC03521 & -17.23 &  9.27 & 7.94 &  6.37$\pm$0.07    & 22.7$\pm$1.3 & 9.17 & 8.84 & 0.39 & 21.4&  37.2\\
NGC4567 & -19.24 & 10.01 &  8.69 & 7.31$\pm$0.07   &  22.7$\pm$1.3 &10.01 & 9.79 & 0.47 & 20.9&  24.0\\
NGC4568 & -19.93 & 10.49 &  8.99 & 7.68$\pm$0.07   &  23.0$\pm$1.4 &10.25 & 10.20 & 0.63 & 31.6&  20.4\\ \hline
\caption{The Herschel Virgo Cluster Survey Bright Galaxy Sample - (1) name, (2) absolute B magnitude, (3) stellar mass, (4) HI mass, (5) dust mass, (6) dust temperature, (7) stellar luminosity from 0.4 to 2.5 $\mu$m, (8) far-infrared luminosity from 100 to 500 $\mu$m, (9) mean optical depth, (10) stellar to atomic gas mass ratio and (11) atomic gas to dust mass ratio. The B, V and H band magnitudes, used to calculate stellar mass, are taken from the GOLDMINE database. The mean fractional dust mass and temperature errors are approximately 19 and 6\% respectively. Seven galaxies have no detected atomic hydrogen. Two galaxies have no dust mass and temperatures listed because their SEDs are not well fitted by a thermal spectrum. Stellar luminosities were calculated using SDSS and 2MASS data. Far-infrared luminosities were calculated using the data from this paper.}
\end{longtable}
\end{center}
\twocolumn
\end{document}